\documentclass[preprint,showpacs,preprintnumbers,amsmath,amssymb]{revtex4}

\usepackage{graphicx}
\usepackage{dcolumn}
\usepackage{bm}
\usepackage{hyperref}
\usepackage{latexsym}
\usepackage{color}

\begin{document}

\title{Percolation of a general network of networks}

\author{Jianxi Gao,$^{1,2}$ Sergey V. Buldyrev,$^3$
H. Eugene Stanley, $^2$ Xiaoming Xu, $^1$ and Shlomo Havlin,$^4$}

\affiliation{$^1$Department of Automation, Shanghai Jiao Tong
University, 800 Dongchuan Road, Shanghai 200240, PR China\\
$^2$Center for Polymer Studies and Department of
Physics, Boston University, Boston, MA 02215 USA\\
$^3$Department of Physics,~Yeshiva University, New York, NY 10033 USA\\
$^4$Department of Physics, Bar-Ilan University, 52900 Ramat-Gan,
Israel}

\date{15 June 2013 --- gbsh061513.tex}

\begin{abstract}
Percolation theory is an approach to study vulnerability of a
system. We develop analytical framework and analyze
percolation properties of a network composed of interdependent
networks (NetONet). Typically, percolation of a single network shows
that the damage in the network due to a failure is a continuous function of the
fraction of failed nodes. In sharp contrast, in NetONet, due to the
cascading failures, the percolation transition may be discontinuous
and even a single node failure may lead to abrupt collapse of the
system. We demonstrate our general framework for a NetONet composed
of $n$ classic Erd\H{o}s-R\'{e}nyi (ER) networks, where each network
depends on the same number $m$ of other networks, i.e., a random
regular network of interdependent ER networks. In contrast to a
\emph{treelike} NetONet in which the size of the largest connected
cluster (mutual component) depends on $n$, the loops in the RR
NetONet cause the largest connected cluster to depend only on $m$.
We also analyzed the extremely vulnerable feedback condition of
coupling. In the case of ER networks, the NetONet only exhibits two phases, a second order
phase transition and collapse, and there is no first phase
transition regime unlike the no feedback condition. In the case of NetONet composed of RR networks, there exists a first order phase transition when $q$ is large and second order phase transition when $q$ is small. Our
results can help in designing robust interdependent systems.
\end{abstract}

\maketitle

\section{Introduction}

Network science has attracted much attention in recent years due to
its interdisciplinary applications
\cite{Strogatz1998,bara2000,Rosato2008,Peerenboom2001,Rinaldi2001,Cohen2000,Callaway2000,Albert2002,Newman2003,
Dorogovtsev2003,song2005,Pastor2006, Caldarelli2007,Barrat2008,
Shlomo2010,Neman2010,
pocock2012robustness,bashan2012network,zhao2012percolation,li2010towards,schneider2011mitigation}.
Many network results have been obtained by analyzing isolated
networks, but most real-world networks do in fact interact with and
depend on other networks
\cite{Rosato2008,Peerenboom2001,Rinaldi2001,bashan2012network,zhao2012percolation}.
Thus, in analogy to the ideal gas laws that are valid only in the
limiting case that molecules do not interact, so the extensive
results for the case of non-interacting networks hold only when it
is justified to neglect the interactions between networks. Recently
several studies have addressed the resilience as well as other
properties of interacting networks
\cite{Sergey2010,parshani2010,huang2011,
dong2012percolation,Shao2011,gao2010,
gao2011pre,gao2012,Alessandro2010, Leicht2011,
morris2012,son2012,saumell2012,gomez2013,aguirre2013,brummitt2012,
Schneider2013, parshani2010epl,yanqing2011,Sergey2011}. A framework
based on percolation theory has been developed to analyze the
cascading failures caused by interdependencies between two networks
\cite{Sergey2010,parshani2010}. In interdependent networks, when
nodes in one network fail they usually cause the failure of
dependent nodes in other networks, and this in turn can cause
further damage to the first network and result in cascading
failures, which could lead to abrupt collapse of the system. Later
on, two important generalizations of the basic
model \cite{Sergey2010,parshani2010} have been developed. Because in
real-world scenarios the initial failure of important nodes
(``hubs'') may not be random but targeted, a mathematical framework
for understanding the robustness of interdependent networks under an
initial targeted attack on specific degree of nodes has been studied
by Huang et al. \cite{huang2011} and later extended by Dong et al.
\cite{dong2012percolation}. Also in real-world scenarios, the
assumption that each node in network A depends on one and only one
node in network B and vice versa may not be valid. To release this
assumption, a theoretical framework for understanding the robustness
of interdependent networks with a random number of support and
dependency relationships has been developed and studied by Shao et
al.\cite{Shao2011}. More recently, Gao et al. developed an
analytical framework to study percolation of a tree-like network
formed by $n$ interdependent networks
\cite{gao2010,gao2011pre,gao2012}. Gao et al. found that while for
$n=1$ the percolation transition is a second order, for any $n>1$
cascading failures occur and the network collapses as in a first
order transition. Indeed cascading failures have caused blackouts in
interdependent communication and power grid systems spanning several
countries \cite{Rosato2008, dobson2007}. To be able to design resilient
infrastructures or improve existing infrastructures we need to
understand how venerability is affected by such interdependencies
\cite{Peerenboom2001,Rinaldi2001,Rosato2008,Alessandro2010,Schneider2013}.

Here we generalize the theory of interdependent networks
\cite{gao2010,gao2011pre,gao2012} to regular and random regular (RR)
network of $n$ interdependent networks that include loops.
Figures~\ref{fig1}(a) and \ref{fig1}(b) illustrate such network of
networks (NetONet), in which each network depends on the same number
$m$ of other networks. We develop an exact analytical approach for
percolation of a regular and a random regular NetONet system
composed of $n$ partially interdependent networks. We show that for
an RR network  with degree $m$ of $n$ interdependent networks where each network has
the same degree distribution, same average degree $<k>$ and the fraction of
dependence nodes between a pair of interdependent networks, $q$, is
the same for all pairs, the number of networks is irrelevant. We
obtain analytically the fraction of survived nodes in each network
after cascading failures, $P_{\infty}$ as a function of $p$, $m$ and $<k>$.

\section{Cascading Failures in a Network of Networks}

\subsection{The Model}
In our model, each node in the NetONet is itself a network and each
link represents a \emph{fully\/} or \emph{partially\/} dependent
pair of networks [see Fig.~\ref{fig1}]. We assume that each network
$i$ ($i=1,2,...,n$) of the NetONet consists of $N_i$ nodes linked
together by connectivity links. Two networks $i$ and $j$ form a
partially dependent pair if a certain fraction $q_{ji}>0$ of nodes
in network $i$ directly depend on nodes in network $j$, i.e., nodes
in network $i$ cannot function if the corresponding nodes in network
$j$ do not function. A node in a network $i$ will not function if it
is removed or if it does not belong to the largest connected cluster
(giant component) in network $i$. Dependent pairs may be connected
by unidirectional dependency links pointing from network $j$ to
network $i$ [see Fig.~\ref{fig1}(c)]. This convention indicates that
nodes in network $i$ may get a crucial support from nodes in network
$j$, e.g., electric power if network $j$ is a power grid.

We assume that after an attack or failure only a fraction of nodes
$p_i$ in each network $i$ remains. We also assume that only nodes
that belong to a giant component in each network $i$ will remain
functional. When a cascade of failures occurs, nodes in network $i$
that do not belong to the giant component in network $i$ fail and
cause nodes in other networks that depend on them to also fail. When those nodes fail, dependent nodes and isolated nodes in the other networks also fail, and the cascade can cause further failures back in network $i$. In order to determine the fraction of nodes
$P_{\infty,i}$ in each network that remains functional (i.e., the
fraction of nodes that constitutes the giant component) after the
cascade of failures as a function of $p_i$ and $q_{ij}$, we need to
analyze the dynamics of the cascading failures.

\subsection{Dynamic Processes}

We assume that all $N_i$ nodes in network $i$ are randomly assigned a degree $k$ from a probability distribution $P_i(k)$, they are randomly connected, and the only constraint is that a node with
degree $k$ has exactly $k$ links \cite{molloy1998}. We define the
generating function of the degree distribution,
\begin{equation}\label{GE1}
G_i(z)\equiv\sum^{\infty}_{k=0}P_i(k)z^k,
\end{equation}
where $z$ is an arbitrary complex variable. The generating function
of this branching process is defined as $H_i(z) \equiv G_i'(z)/
G_i'(1)$. Once a fraction $1-x$ of nodes is randomly removed from a
network, the probability that a randomly chosen node belongs to a
giant component, is given by
\cite{Sergey2010,parshani2010,Newman2001,Newman2002PRE,Shao2008,Shao2009}
\begin{equation}\label{GE2}
g_i(x)=1-G_i[xf_i(x)+1-x],
\end{equation}
where $f_i(x)$ satisfies
\begin{equation}\label{GE3}
f_i(x)=H_i[xf_i(x)+1-x].
\end{equation}

We assume that (i) each node $a$ in network $i$ depends with a
probability $q_{ji}$ on only one node $b$ in network $j$, and that,
(ii) if node $a$ in network $i$ depends on node $b$ in network $j$
and node $b$ in network $j$ depends on node $c$ in network $i$, then
node $a$ coincides with node $c$, i.e., we have a no-feedback
situation \cite{gao2012}. In section IV we study the case of
feedback condition, i.e., node $a$ can be different from $c$ in
network $i$. The no feedback condition prevents configurations from
collapsing even without having their internal connectivity in each
network \cite{Shao2011}. Next, we develop the dynamic process of
cascading failures step by step.

At $t=1$, in networks $i$ of the NetONet we randomly remove a
fraction $1-p_i$ of nodes. After the initial removal of nodes, the
remaining fraction of nodes in network $i$, is $\psi'_{i,1} \equiv
p$. The remaining functional part of network $i$ therefore
constituents a fraction $\psi_{i,1}=\psi'_{i,1}g_A(\psi'_{i,1})$ of
the network nodes, where $g_i(\psi'_{i,1})$ is defined by
Eqs.~(\ref{GE2}) and (\ref{GE3}). Furthermore, we denote by
$y_{ji,1}$ the fraction of nodes in network $i$ that survive after
the damage from all the networks connected to network $i$ except
network $j$ is taken into account, so if $q_{ij} \neq 0$, $y_{ji,1}
= p_j$.

When $t \geq 2$, all the networks receive the damages from their
neighboring networks one by one. Without loss of generality, we
assume that network $1$ is the first, network $2$ second,..., and
network $n$ is last. In Fig. 1(c), for example, since a fraction
$q_{21}$, $q_{31}$, $q_{41}$ and $q_{71}$ of nodes of network $1$
depends on nodes from network $2$, $3$, $4$ and $7$ respectively,
the remaining fraction of network $1$ nodes is,

\begin{equation}\label{GE4}
\psi'_{1,t} = \prod_{j=2,3,4,7}
[q_{j1}y_{j1,t-1}g_j(\psi'_{j,t-1})-q_{j1}+1],
\end{equation}
and $y_{1j,t}$ ($j=2,3,4,7$) satisfies

\begin{equation}\label{GE6}
y_{1j,t} =
\frac{\psi'_{1,t}}{q_{j1}y_{j1,t-1}g_j(\psi'_{j,t-1})-q_{j1}+1}.
\end{equation}
The remaining functional part of network $1$ therefore contains a
fraction $\psi_{1,t}=\psi'_{1,t}g_1(\psi'_{1,t})$ of the network
nodes.

Similarly, we obtain the remaining fraction of network $i$ nodes,

\begin{equation}\label{GE7}
\psi'_{i,t} = \prod_{j<i}
[q_{ji}y_{ji,t-1}g_j(\psi'_{j,t})-q_{ji}+1]\prod_{s>i}
[q_{si}y_{si,t-1}g_s(\psi'_{s,t-1})-q_{si}+1],
\end{equation}

and $y_{ij,t}$ is
\begin{equation}\label{GE8}
y_{ij,t} =
\frac{\psi'_{i,t}}{q_{ji}y_{ji,t-1}g_j(\psi'_{j,t})-q_{ji}+1},
\end{equation}
and $y_{is,t}$ is
\begin{equation}\label{GE9}
y_{is,t}=\frac{\psi'_{i,t}}{q_{si}y_{si,t-1}g_s(\psi'_{s,t-1})-q_{si}+1}.
\end{equation}

Following this approach we can construct the sequence, $\psi'_{i,t}$
of the remaining fraction of nodes at each stage of the cascade of
failures. The general form is given by
\begin{equation}\label{GE10}
\begin{array}{lcl}
\psi'_{i,1} \equiv p_i, & \mbox{} & \\
y_{ij,1} \equiv p_i, q_{ij} \neq 0 & \mbox{} &  \\
\psi'_{i,t} = p_i\prod_{j<i}{[q_{ji}y_{ji,t-1}g_j(\psi'_{j,t})-q_{ji}+1]}\prod_{s>i}{[q_{si}y_{si,t-1}g_s(\psi'_{s,t-1})-q_{si}+1]}, & \mbox{} & \\
y_{ij,t}=\frac{\psi'_{i,t}}{q_{ji}y_{ji,t-1}g_j(\psi'_{j,t})-q_{js}+1}, & \mbox{} & \\
y_{is,t}=\frac{\psi'_{i,t}}{q_{si}y_{si,t-1}g_s(\psi'_{s,t-1})-q_{si}+1}. & \mbox{} &\\
\end{array}
\end{equation}

We compare the theoretical formulas of the dynamics,
Eqs.~(\ref{GE10}) and simulation results in Fig.~\ref{fig2}. As seen
the theory of the dynamics (\ref{GE10}) agrees well with
simulations.

\subsection{Stationary State}
To determine the state of the system at the end of the cascade
process we look at $\psi'_{i,\tau}$ at the limit of $\tau\rightarrow
\infty$. This limit must satisfy the equations
$\psi'_{i,\tau}=\psi'_{i,\tau+1}$ since eventually the clusters stop
fragmenting and the fractions of randomly removed nodes at step
$\tau$ and $\tau+1$ are equal. Denoting $\psi'_{i,\tau}=x_i$, we
arrive for the $n$ networks, at the stationary state, to a system of
$n$ equations with $n$ unknowns,

\begin{equation}\label{E1}
x_i=p_{i} \prod_{j=1}^{K}{(q_{ji}y_{ji}g_j(x_j)-q_{ji}+1)},
\end{equation}
where the product is taken over $K$ networks interlinked with
network $i$ by partial (or fully) dependency links [see
Fig.~\ref{fig1}] and
\begin{equation}\label{E2}
y_{ij}=\frac{x_i}{q_{ji}y_{ji}g_j(x_j)-q_{ji}+1},
\end{equation}
is the fraction of nodes in network $i$ that survive after the
damage from all the networks connected to network $j$ except network
$i$ itself is taken into account. The damage from network $i$ itself is
excluded due to the no-feedback condition. Equation~(\ref{E1}) is
valid for any type of interdependent NetONet, while Eqs.~(\ref{E2})
represents the no-feedback condition. For two coupled networks,
Eqs.~(\ref{E1}) and (\ref{E2}) are equivalent to Eq.~(13) of
Ref.~\cite{Shao2011} for the specific case of single dependency
links.

Our general framework for percolation of interdependent network of
networks, Eqs.~(\ref{E1}) and (\ref{E2}), can be generalized in two
directions: (i) coupling with feedback condition (ii) coupling with
multiple-support.

(i) In the existence of the feedback, $y_{i,j}$ is simply $x_i$ and
Eqs.~(\ref{E1}) and (\ref{E2}) become a single equation,

\begin{equation}\label{E1-1}
x_i=p_{i} \prod_{j=1}^{K}{(q_{ji}x_jg_j(x_j)-q_{ji}+1)}.
\end{equation}
The feedback condition leads to an extreme vulnerability of the network of
interdependent networks. As we know for two fully interdependent
networks with no-feedback condition \cite{Sergey2010} if the average
degree is large enough both networks exist. However, for two fully
interdependent networks with feedback condition, no matter how large
the average degree is, both networks collapse even after a single node is removed. The analytical
results about the feedback condition are given in section IV.

(ii) Equation~(\ref{E1}) can be generalized to the case of multiple
dependency links studied for a pair of coupled networks in \cite{Shao2011} by,

\begin{equation}\label{E1-2}
x_i =p_i\prod_{j=1}^K \left( 1-q_{ji}G^{ji}[1-x_jg_j(x_j)]\right),
\end{equation}
where $G^{ji}$ represents the generating function of the degree
distribution of multiple support links that network $i$ depends on
network $j$.

On one hand, the term $g_i$ reflects the topology of network $i$,
which can be an ER network, a RR network, a scale free (SF) network,
or even a small world (SW) network. On the other hand,
$Q=[q_{ij}]_{n\times n}$ ($n$ is the number of networks) reflects
the interactions between the networks, i.e., the topology of the
NetONet, which can also be any type of network. Our theoretical
results Eq.~(\ref{E1}) and (\ref{E2}) are therefore general for any
type of network of networks. By solving Eqs.~(\ref{E1}) and
(\ref{E2}), or Eqs.~(\ref{E1-1}) , or Eqs. ~(\ref{E1-2}), we obtain
$x_i$ of each network for coupled networks with no feedback
condition, feedback condition and multiple-support condition,
respectively. Thus, we obtain the giant component in each network
$i$ as

\begin{equation}\label{E1-3}
P_{\infty,i}\equiv x_ig_i(x_i).
\end{equation}
\section{No Feedback condition}

\subsection{The general case of an RR NetONet formed of random networks.}

In order to study the various forms the stationary state of the system can reach
after a cascading failure, we first assume, without loss of
generality, that each network depends on $m$ other random networks,
i.e., that we have a RR network formed of $n$ random networks. We
understand the RR category to also include regular non-random
networks in which each network has the same number of neighbouring
interdependent networks with a structure e.g., of a lattice of ER
networks [Fig.~\ref{fig1}(a)]. We assume, for simplicity, that the
initial attack on each network is by removing randomly a fraction
$1-p$ of nodes, the partial interdependency fraction is $q$, and the
average degree of each ER network is the same $\bar{k}$ for all
networks. Because of the symmetries involved, the $nm+m$ equations
in Eqs.~(\ref{E1}) and (\ref{E2}) can be reduced to two equations,
\begin{equation}\label{E11}
\left\{
\begin{array}{lcl}
x = p(qyg(x)-q+1)^m, & \mbox{} &  \\
y = p(qyg(x)-q+1)^{m-1}. & \mbox{} &  \\
\end{array}\right.
\end{equation}

By substituting $z=xf(x)+1-x$,  Eqs. (\ref{GE2}), (\ref{GE3}) and (\ref{E1-3}) into (\ref{E11}), and eliminating $f$, $x$, and $y$, we
obtain

\begin{equation}\label{GE12}
P_{\infty}(z)=\frac{[1-G(z)](1-z)}{1-H(z)},
\end{equation}
and

\begin{equation}\label{GE5}
\left(\frac{1-z}{1-H(z)}\right)^{\frac{1}{m}}\left(\frac{1}{p}\right)^{\frac{2}{m}}+
(q-1)\left(\frac{1}{p}\right)^{\frac{1}{m}}-qP_{\infty}(z)\left(\frac{1-H(z)}{1-z}\right)^{\frac{1}{m}}=0.
\end{equation}

Equation (\ref{GE5}) can help us to understand the percolation of a RR
network of any interdependent random networks where all networks have the same average degree and degree distribution.

To solve Eq. (\ref{GE5}), we introduce an analytical function $R(z)$
for $z\in[0,1]$ as

\begin{equation}\label{GE12-1}
\frac{1}{p}=\frac{H(z)-1}{z-1}\left(\frac{1-q+\sqrt{(1-q)^2+4qP_{\infty}(z)}}{2}\right)^m
\equiv R(z).
\end{equation}

$R(z)$ as a function of $z$ has a quite complex behaviour for
various degree distributions. We present two examples to demonstrate
our general results on (i) RR network of ER networks and (ii) RR
network of SF networks.

\begin{itemize}

\item[{(i)}] For the case of RR network of ER networks we find a critical $q_c$ such that, when $q<q_c$ the system shows a second order phase transition and the critical
threshold $p_c$ depends on $q$ and average degree $\bar{k}$. When
$q_c<q<q_{\max}$ the system shows a first order phase transition,
and when $q>q_{\max}$ there is no phase transition because all the
networks collapse even for a single node failure.

\item[{(ii)}] For the case of RR network of SF networks, the phase diagram is different from the ER case, because there is no
pure first order phase transition. However, there exists an
effective $q^e_c$, when $q<q^e_c$, the system shows a second order
phase transition and the critical threshold is $p_c=0$ for infinite
number of nodes in each network, i.e., the maximum degree goes to
$\infty$. When $q^e_c<q<q_{\max}$, the system shows a hybrid
transition as follows. When $p$ decreases from 1 to 0, the giant
component $P_{\infty}$ as function of $p$ shows a sharp jump at
$p^{I}_{ec}$, which is like a first order transition to a finite
small value, and then (when $p$ further decreases) goes smoothly to
0. For $q>q_{\max}$ there is no phase transition because all the
networks collapse even for a single node failure.
\end{itemize}

\subsection{RR network formed by interdependent ER networks}

For ER networks \cite{ER1959,ER1960,Bollob1985}, the generating
function $g(x)$
satisfies~\cite{Newman2001,Newman2002PRE,Shao2008,Shao2009}

\begin{equation}\label{E13-0}
\begin{array}{lcl}
g(x) =1-\exp[\bar{k}x(f-1)], & \mbox{} &  \\
f=\exp[\bar{k} x(f-1)].
\end{array}
\end{equation}

Substituting Eqs.~(\ref{E13-0}) into Eqs.~(\ref{E11}), we get
\begin{equation}\label{E13}
\begin{array}{lcl}
f=\exp{\{\bar{k}p[qy(1-f)-q+1]^m(f-1)\}}, & \mbox{} &  \\
y = p[qy(1-f)-q+1]^{m-1}, & \mbox{} &  \\
P_{\infty} = -(\log f)/\bar{k}.
\end{array}
\end{equation}
Eliminating $y$ from Eq.~(\ref{E13}), we obtain an equation for $f$,
\begin{equation}\label{E14}
\begin{array}{lcl}
[\frac{\ln f}{\bar{k}p(f-1)}]^{\frac{2}{m}}+(q-1)[\frac{\ln
f}{\bar{k}p(f-1)}]^{\frac{1}{m}}+\frac{q}{\bar{k}}\log f =0.
\end{array}
\end{equation}
Considering $[\ln f/\bar{k}p(f-1)]^{1/m}$ to be a variable, Eq.~(\ref{E14})
becomes a quadratic equation that can be solved analytically having
only one valid solution,
\begin{equation}\label{E15}
2^m\ln f =
\bar{k}p(f-1)\left[1-q+\sqrt{(1-q)^2-\frac{4q}{\bar{k}}\ln
f}\right]^m.
\end{equation}
From Eq.~(\ref{E15}) and the last equation in (\ref{E13}), we determine
the mutual percolation giant component for a RR network of ER networks,
\begin{equation}\label{E16}
P_{\infty}=\frac{p}{2^m}(1-e^{-\bar{k}P_{\infty}})
\left[1-q+\sqrt{(1-q)^2+4qP_{\infty}}\right]^m.
\end{equation}

Figures \ref{fig3}(a) and \ref{fig3}(b) show numerical solutions of
Eq.~(\ref{E16}) for several $q$ and $m$ values compared with
simulations. These solutions imply that $P_{\infty}$ as a function
of $p$ exhibits a second (continuous) or a first order (abrupt)
phase transition depending on the values of $q$ and $m$ for a given
$\bar{k}$. Note, when q=0 or m=0, Eq. (\ref{E16}) is reduced to the
known equation, $P_{\infty}=p(1-e^{-\bar{k}P_{\infty}})$, for single
ER networks \cite{ER1959,ER1960,Bollob1985}.

From Eqs. (\ref{GE12-1}) and (\ref{E16}), we obtain

\begin{equation}\label{GE15}
R(z)=\frac{1}{p}=\frac{(1-e^{\bar{k}(z-1)})[1-q+\sqrt{(1-q)^2+4q(1-z)}]^m}{2^m(1-z)},
\end{equation}
and
\begin{equation}\label{GE16}
\begin{split}
F(z)\equiv\frac{ \rm{d} R(z)}{\rm{d} z} & =\frac{e^{\bar{k}(1-z)}-\bar{k}(1-z)}{p(1-z)(e^{\bar{k}(1-z)}-1)}\\
& -\frac{2mq}{p[1-q+\sqrt{(1-q)^2+4q(1-z)}]\sqrt{(1-q)^2+4q(1-z)}}.
\end{split}
\end{equation}

Next we demonstrate the behaviour of Eq.~(\ref{GE15}) as shown in
Fig.~\ref{fig4}. For given $\bar{k}$ and $m$, when $q$ is small,
$R(z)$ is a monotonously increasing function of $z$, for example see
the curve for $q=0.42$. Thus, the maximum of $R(z)$ is obtained when
$z\rightarrow1$, which corresponds to a second order phase
transition threshold $p^{II}_c = 1/\max\{R(z)\} \equiv 1/R(z_c)$,
where $P_{\infty}(p^{II}_c) = 1-z_c=0$. When $q$ increases, $R(z)$
as a function of $z$ shows a maxima at $z<1$ and $\max\{R(z)\}>1$,
for example for $q=0.50$ in Fig.~\ref{fig4}. Thus, the maximum of
$R(z)$ is obtained when $z = z_c \in (0,1)$ at the peak, which
corresponds to the first order phase transition threshold $p^{I}_c =
1/\max\{R(z)\} = 1/R(z_c)$, where $P_{\infty}(p^{I}_c) = 1-z_c > 0$.
The $q$ value in which for the first time a maxima of $R(z)$ appears at
$z<1$ is $q_c$, the critical dependency which separates between the
first and second order transitions. When $q$ continually further
increases, $\max\{R(z)\} < 1$, which corresponds to a complete
collapse of the NetONet. The value of $q$ for which $\max\{R(z)\}=1$
is $q_{max}$, above which the network is not stable and collapse
instantaneously.

Next we analyze the different behaviours of RR network of ER
networks in the different regimes of $q$: (i) For $q<q_c$, the
percolation is a continuous second order which is characterized by a
critical threshold $p^{II}_c$. (ii) The range of $q_c<q<q_{\max}$ is
characterized by an abrupt first order phase transition with a
critical threshold $p^{I}_c$. (iii) For $q>q_{\max}$ no transition
exists due to the instant collapse of the system.

We next analyze in detail the parameters characterizing the three
regimes. (i) For a given $m$ and $\bar{k}$, when $q$ is sufficiently
small, there exists a critical $p^{II}_c$ such that, when $p$
increases above $p^{II}_c$, $P_{\infty}$ continuously increases from
zero to non-zero values. Here $P_{\infty}$ as a function of $p$
exhibits a second order phase transition. In order to get $p^{II}_c$
we analyze Eq.~(\ref{GE16}). When $q$ is sufficiently small $\frac{
\rm{d} R(z)}{\rm{d} z}>0$, the maximum value of $R(z)$ is obtained
when $z\rightarrow 1$. Thus, we obtain the critical threshold for
the second order phase transition, $p^{II}_c$ by substituting
$z\rightarrow 1$ into Eq.~(\ref{GE15}),

\begin{equation}\label{E20}
p^{II}_c  = \frac{1}{\bar{k}(1-q)^m}.
\end{equation}

(ii) Next we obtain $p_c^I$. According to Eq.~(\ref{GE16}), when $q$
increases, $R(z)$ as a function $z$ becomes not monotonous and a
maxima appears, which corresponds to the condition for first order
phase transition, i.e., when $\frac{ \rm{d} R(z)}{\rm{d} z} = 0$.
Furthermore, for a given $p$, the smallest of these roots gives the
physically meaningful solution from which the giant component
$0<P_{\infty}(p_c)<1$ can be found from Eq.~(\ref{E16}).

By solving $z_c$ from $F(z_c)=0$ of Eq.~(\ref{GE16}), we obtain the
critical threshold for first order phase transition $p^{I}_c$ as

\begin{equation}\label{GE18}
p^{I}_c=\frac{2^m(1-z_c)}{(1-e^{\bar{k}(z_c-1)})[1-q+\sqrt{(1-q)^2+4q(1-z_c)}]^m}.
\end{equation}

Next we study the critical coupling strength $q_c$, i.e, the
critical coupling that distinguishes between first and second order
transitions. We find that $P_{\infty}$ undergoes a second order
transition as a function of $p$ when $q < q_c$, a first order
transition when $q_c<q<q_{\max}$, and no phase transition (the
system is unstable for any $p$) when $q>q_{\max}$. By definition,
when a system changes from second order to first order at the
critical point, $q$, $m$, and $\bar{k}$ satisfy $p^{I}_c=p^{II}_c$,
i.e., both conditions for the first order and second order phase
transition should satisfy,

\begin{equation}\label{GE19}
\lim_{z\rightarrow1}\frac{ \rm{d} R(z)}{\rm{d} z} = 0.
\end{equation}

From Eqs. (\ref{GE16}) and (\ref{GE19}), we obtain
\begin{equation}\label{E24}
2qm-\bar{k}(1-q)^2=0.
\end{equation}
Solving Eq. (\ref{E24}), we find that the physically meaningful
$q_c$ is
\begin{equation}\label{E25}
q_c = \frac{\bar{k}+m-({m^2+2\bar{k}m})^{1/2}}{\bar{k}}.
\end{equation}

(iii) Next we calculate the critical point $q_{\max}$, above which
($q>q_{\max}$) the system is unstable for any $p$. From Eq.
(\ref{GE15}), the system is unstable for any $p$, when $R(z) \leq
1$. We therefore, can obtain $q_{\max}$ by satisfying Eq.
(\ref{GE16}) and $p^{I}_c=1$. Thus, we obtain $q_{\max}$ as

\begin{equation}\label{E25-1}
q_{\max} = \frac{(a^{1/m}-1)^2}{2(1-2z_c-a^{1/m})},
\end{equation}
where $a$ satisfies

\begin{equation}\label{E25-2}
a = \frac{1-e^{\bar{k}(z_c-1)}}{2^m(1-z_c)},
\end{equation}
and $z_c$ can be solved by substituting Eq. (\ref{E25-1}) into
Eq.~(\ref{GE16}) and set $p=1$, $F(z_c)=0$, which is one equation
with only one unknown $z_c$.

Next we obtain the numerical solution of $P_{\infty}(p_c)$ as a
function $q$ as shown in Fig. \ref{fig5}. From Fig. \ref{fig5}, we
can see that for fixed $m$ and $\bar{k}$, there exist two critical
values of coupling strength, $q_c$ and $q_{\max}$, when $q<q_c$,
$P_{\infty}(p_c)=0$ which represents a second order phase
transition, when $q_c<q<q_{\max}$, $P_{\infty}(p_c)>0$ representing
a first order phase transition. When $q>q_{\max}$,
$P_{\infty}(p_c)=0$ representing the NetONet collapse and that there
is no phase transition ($p_c=1$). Figure \ref{fig5-1} shows the
phase diagram of RR network of ER networks for different values of
$m$ and $\bar{k}$. As $m$ decreases and $k$ increases, the region
for $P_{\infty}>0$ increases, which shows a better robustness.

\subsection{The case of RR NetONet formed of interdependent scale-free (SF) networks.}

We analyze here NetONets composed of SF networks with a power law
degree distribution $P(k)\sim k^{-\lambda}$. The corresponding
generating function is
\begin{equation}\label{e26}
G(z) =
\frac{\sum^M_s[(k+1)^{1-\lambda}-k^{1-\lambda}]z^k}{(M+1)^{1-\lambda}
-s^{1-\lambda}}
\end{equation}
where $s$ ($s=2$ in this paper) is the minimal degree cutoff and $M$ is the maximal degree cutoff.

SF networks approximate real networks such as the Internet, airline
flight patterns, and patterns of scientific collaboration
\cite{Barabasi1999,colizza2006,Albert2002,lidaqing2011}. When SF
networks are fully interdependent \cite{Sergey2010}, $p_c>0$, even
in the case $\lambda\leq 3$ in contrast to a single network for
which $p_c=0$ \cite{Cohen2000}. We study the percolation of a RR
network composed of interdependent SF networks by substituting their
degree distribution into Eq.~(\ref{GE1}) and obtaining their
generating functions. We assume, for simplicity, that all the
networks in the NetONet have the same $\lambda$, $s$ and $M$, and use
Eq.~(\ref{GE5}) to analyze the percolation of an RR NetONet of SF
networks.

The generating function of the branching process is defined as
$H(z)=G'(z)/ G'(1)$. Substituting $H(z)$ and Eq. (\ref{e26}) into
Eq. (\ref{GE12-1}), we obtain the function $R(z)$ for RR of SF
networks. As shown in Fig. \ref{fig5-20}, we find three regimes of
coupling strength $q$:

\begin{itemize}

\item[{(i)}] When $q$ is small
($q<q^e_c$), $R(z)$ is a monotonically increasing function of $z$,
the system shows a second order phase transition, and the critical
threshold $p^{II}_{c}$ is obtained when $z\rightarrow1$ which
corresponds to $R(1)=\max\{R\}=\infty=1/p^{II}_c$, i.e.,
$p^{II}_c=0$.

\item[{(ii)}] When $q$ is larger, $q^e_c<q<q_{\max}$, $R(z)$ as a
function of $z$ shows a peak which corresponds to a sharp jump to a
lower value of $P_\infty$ at $z_c$ with a hybrid transition, because
$\max\{R\} \neq R(z_c)$, which is different from the ER case.
Furthermore, the effective critical threshold (sharp jump) is
$p^{I}_{ec}=1/R(z_c)$, while for $p$ below this sharp jump the
system undergoes a smooth second order phase transition and the
critical threshold is zero, similar to (i). Thus, when $z$ is
greater than some value, $R(z)$ increases with $z$ again and reaches
$\max\{R\}$ when $z\rightarrow1$, which indicates that when $p$
decreases below $p^{I}_{ec}$, $P_{\infty}$ jumps as a first order to
a finite small value and then decreases smoothly to 0 as $p$
approaches $0$;

\item[{(iii)}] When $q$ is above $q_{\max}$, $R(z)$ decreases with
$z$ first, and then increases with $z$, which corresponds to the
system collapse.
\end{itemize}

Next we analyze the three regimes more rigorously.

(i) When $q$ is small ($q<q_c^e$), $R(z)$ is a monotonically
increasing function of $z$, the maximum of $R(z_c)$ is obtained when
$z_c\rightarrow1$, which corresponds to $P_{\infty}=0$,
\begin{equation}\label{e27}
\max\{R\}=\lim_{z\rightarrow1}\frac{H(z)-1}{z-1}\left(1-q\right)^m\doteq
H^\prime (1).
\end{equation}
This is since when $M\rightarrow \infty$, $\max\{R\}\rightarrow
\infty$, $p^{II}_c=0$ when $q<q^e_c$.

(ii) As $q$ increases ($q \geq q^e_c$), $R(z)$ as a function of $z$
shows a peak corresponding to $R(z) = R(z_c)$, $\rm{d}R/\rm{d}z=0$
(smaller root has the physical meaning), where
$R=R_c=1/p^{I}_{ec}>1$ corresponding to the effective critical
threshold where $P_{\infty}$ as a function of $p$ shows an abrupt
jump. Furthermore, we define

\begin{equation}\label{e27-0}
P^-_{\infty}=\lim_{p\rightarrow p^{I}_{ec},p<p^I_{ec}}P_{\infty}(p),
\end{equation}
and
\begin{equation}\label{e27-00}
P^+_{\infty}=\lim_{p\rightarrow p^{I}_{ec},p>p^I_{ec}}P_{\infty}(p).
\end{equation}
For the case of first order phase transition with a sharp jump,
$P^-_{\infty}=0$, but for the hybrid transition $P^-_{\infty}>0$.
After the sharp jump, $P_{\infty}$ decreases smoothly to 0 until
$p=0$. For the case of two partially interdependent SF networks see Zhou et.
al. \cite{Zhou2013}.

(iii) As $q$ increases further ($q>q_{max}$), $\frac{ \rm{d}
R(z)}{\rm{d} z}$ at $z=0$ becomes negative, thus the NetONet will
collapse even when a single node is initially removed. So the maximum
values of $q$ is obtained as
\begin{equation}\label{e27-1}
\frac{ \rm{d} R(z)}{\rm{d} z}|_{z\rightarrow 0} =0.
\end{equation}

Using Eqs. (\ref{GE12}), (\ref{GE12-1}) and (\ref{e27-1}), we obtain

\begin{equation}\label{e27-2}
\begin{split}
\frac{ \rm{d} R(z)}{\rm{d} z}
&=-\frac{G^\prime(z)R(z)}{1-G(z)}-\frac{R(z)P^\prime_{\infty}(z)}{P_{\infty}(z)}\\
&+\frac{2mR(z)}{1-q+\sqrt{(1-q)^2+4qP_{\infty}(z)}}\frac{qP^\prime_{\infty}(z)}{\sqrt{(1-q)^2+4qP_{\infty}(z)}}
\end{split}
\end{equation}
When $q=q_{\max}$, $P_{\infty}(z)|_{z\rightarrow 0}=1$ and
$P^\prime_{\infty}(z)|_{z\rightarrow 0}=-1$, so we get

\begin{equation}\label{e27-3}
q_{\max}=\frac{1}{m-1}.
\end{equation}

Comparison between analytical and simulation results are shown in
Fig.~\ref{fig5-21}.

\section{Feedback condition}

The above detailed analysis considers the case of no feedback
condition since even for two fully interdependent networks with
feedback condition (fb), both networks will completely collapse even
if a single node fails. However, feedback condition can not destroy
a network of partially interdependent networks when $q$ is
sufficiently small. For the case of feedback condition,
Eqs.~(\ref{E11}) become

\begin{equation}\label{GE20}
x = p(qxg(x)-q+1)^m.
\end{equation}

Substituting $z=xf(x)+1-x$ and Eqs.~(\ref{GE1})-(\ref{GE3}) into
Eq.~(\ref{GE20}) and eliminating $x$, we obtain

\begin{equation}\label{GE20-1}
\frac{1}{p}=\frac{1-H(z)}{1-z}(1-q+qP_{\infty})^m
\end{equation}

For ER networks, we obtain an equation for $f$
\begin{equation}\label{GE21}
\ln f = \bar{k}p(1-q-q\frac{\ln f}{\bar{k}})^{-m}(f-1).
\end{equation}
By substituting $P_{\infty} = -(\log f)/\bar{k}$, we determine the
mutual percolation giant component for a RR network of ER networks
with feedback condition,
\begin{equation}\label{GE22}
P_{\infty}=p(1-e^{-\bar{k}P_{\infty}})(1-q+qP_{\infty})^m.
\end{equation}

Figure \ref{fig7} shows numerical solutions of Eq.~(\ref{GE22})
for several $q$ and $m$ values, which are in excellent agreement
with simulations, presented as symbols. These solutions imply that $P_{\infty}$ as a
function of $p$ exhibits only a second order phase transition.

Indeed, from Eq. (\ref{GE22}) and substituting $P_{\infty}=z$
($z\in[0,1]$), we obtain

\begin{equation}\label{GE23}
R(z)=\frac{1}{p}=\frac{(1-e^{-\bar{k}z})}{z}(1-q+qz)^m,
\end{equation}
and
\begin{equation}\label{GE24}
\frac{ \rm{d} R(z)}{\rm{d}
z}=\frac{\bar{k}z-e^{\bar{k}z}+1}{pz(e^{\bar{k}z}-1)}-\frac{mq}{p(1-q+qz)}.
\end{equation}

Next, we prove that $R(z)$ is a decreasing function of $z$, i.e., $
\rm{d} R(z)/\rm{d} z<0$. It is easy to see

\begin{equation}\label{GE24}
\frac{ \rm{d} }{\rm{d}
z}(\bar{k}z-e^{\bar{k}z}+1)=\bar{k}-\bar{k}e^{\bar{k}z} \leq 0,
\end{equation}
and the equal condition is satisfied only when $z=0$, so
$\bar{k}z-e^{\bar{k}z}+1<0$. Thus we obtain that $R(z)$ is a
monotonous decreasing function of $z$, which is very different from
the no feedback condition. So the maximum of $R(z)$ is obtained only
when $z\rightarrow 0$, which corresponds to the critical value of
$p_c$,

\begin{equation}\label{GE25}
p_c  = \frac{1}{\bar{k}(1-q)^m},
\end{equation}
which is the same as Eq. (\ref{E20}). Thus, the second order
threshold of no feedback $p^{II}_c$ is the same as the feedback
$p_c$, which is also shown in Fig. \ref{fig9} (a).  However, the
feedback case is still more vulnerable than the no feedback case.
Fig. \ref{fig9} (b) and (c) show $P_{\infty}$ for $p=1$, i.e. the
giant component in each network of the NetONet when there is no node
failures, as a function of $q$. We can see that for the no feedback
case, Fig. \ref{fig9} (b), the system still has very large giant
component left when both $m$ and $q$ are large, but for the feedback
case, there is not giant component when both $m$ and $q$ are large.
This happens because of the single connected nodes and isolated
nodes in each network \cite{gao2011pre}.

Substituting $p_c \leq 1$ into Eq.~(\ref{GE25}), we obtain $\bar{k}
\geq 1/(1-q)^m$ or $q \leq 1-(1/\bar{k})^{1/m}$, which represents
the minimum $\bar{k}$ and maximum $q$ for which a phase transition
exists,
\begin{equation}\label{GE26}
\bar{k}_{\min} = \frac{1}{(1-q)^m},
\end{equation}
and
\begin{equation}\label{GE27}
q_{\max}=1-(1/\bar{k})^{1/m}.
\end{equation}
Equations (\ref{GE26}) and (\ref{GE27}) demonstrate that the NetONet
collapses when $q$ and $m$ are fixed and $\bar{k} < \bar{k}_{\min}$
and when $m$ and $\bar{k}$ are fixed and $q > q_{\max}$, i.e., there
is no phase transition in these zones. However, $q_{\max}$ of the feedback case is smaller than that of no feedback
case shown in Fig. \ref{fig8} (a), which shows that the feedback
case is more vulnerable than the no feedback case. In Fig. \ref{fig8} (b) we show that increasing
$\bar{k}$ or decreasing $m$ will increase $q_{\max}$, i.e., increase
the robustness of NetONet.

Next we study the feedback condition for the case of RR NetONet formed of RR networks of degree $k$. In this case, Eq. \ref{GE22} becomes 

\begin{equation}\label{GE28}
1-\left[1-\frac{P_{\infty}}{p(1-q-qP_{\infty})}\right]^{\frac{1}{k}}=p\left\{1-\left[1-\frac{P_{\infty}}{p(1-q-qP_{\infty})}\right]^{\frac{k-1}{k}}\right\} (1-q+qP_{\infty})^m.
\end{equation}

We find that the RR networks are very different from the ER networks, and the system shows first order phase transition for large $q$ and a second order phase transition for small $q$ as shown in Fig.  \ref{fig12}.
\section{Discussion}

In summary, we develop a general framework, Eqs.~(\ref{E1}) and
(\ref{E2}), for studying percolation in several types of NetONet of
any degree distribution. We demonstrate our approach for a RR
network of ER networks that can be exactly solved analytically,
Eqs.~(\ref{E16}) and for RR of SF networks for which the analytical
expressions can be solved numerically. We find that $q_{\max}$ and
$q^e_c$ exist, where a NetONet shows a second-order transition when
$q<q^e_c$, a hybrid transition when $q^e_c<q<q_{\max}$, and that in
all other cases there is no phase transition because all nodes in
the NetONet spontaneously collapse. Thus the percolation theory of a
single network is a limiting case of a more general case of
percolation of interdependent networks. Our results show that the
percolation threshold and the giant component depend solely on the
average degree of the ER network and the degree of the RR network,
but not on the number of networks. These findings enable us to study
the percolation of different topologies of NetONet. We expect this
work to provide insights leading to further analysis of real data on
interdependent networks. The benchmark models we present here can be
used to study the structural, functional, and robustness properties
of interdependent networks. Because, in real NetONets, individual
networks are not randomly connected and their interdependent nodes
are not selected at random, it is crucial that we understand many
types of correlations existing in real-world systems and to further
develop the theoretical tools studying all of them. Future studies
of interdependent networks will need to focus on (i) an analysis of
real data from many different interdependent systems and (ii) the
development of mathematical tools for studying the vulnerability of
real-world interdependent systems.

\begin{figure}[h!]
\centering\vspace{0.1in}
\includegraphics[width=0.4\textwidth]{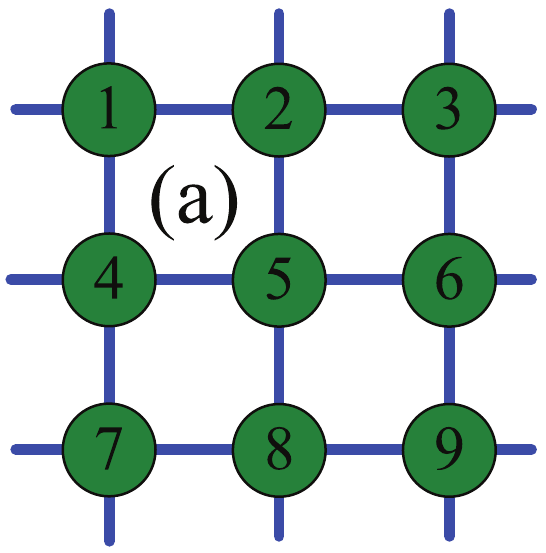}\hspace{0.5in}
\includegraphics[width=0.42\textwidth]{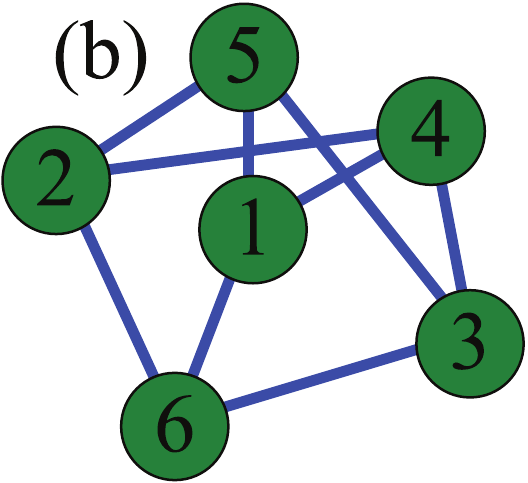}\hspace{0.5in}
\vspace{0.1in}
\includegraphics[width=0.42\textwidth]{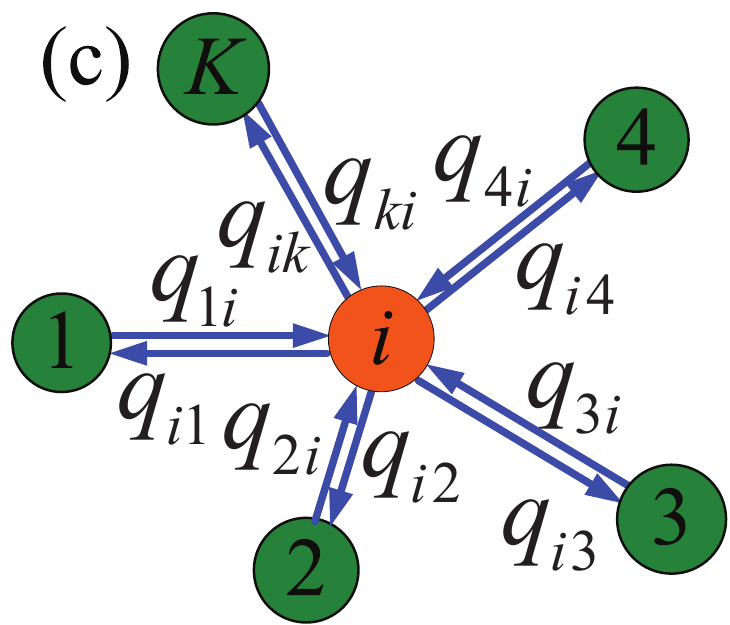}\hspace{0.5in}
\caption{\textbf{Illustration of regular and random regular (RR)
NetONet of interdependent random networks.} (a) An example of a
regular network, a lattice with periodic boundary condition composed
of 9 interdependent networks represented by 9 circles.  The degree
of the NetONet is $m=4$, i.e., each network
  depends on 4 networks. (b) A RR network composed of 6 interdependent
  networks represented by 6 circles. The degree of the NetONet is $m=3$,
  i.e., each network depends on 3 networks. The analytical results for the
  NetONet [Eqs. (\ref{E11}) and (\ref{GE5})] are exact and the same for both cases (a) and (b).
  (c) Schematic representation of the dependencies of the networks. Circles represent
 networks in the NetONet, and the arrows represent the partially
  interdependent pairs. For example, a fraction of $q_{3i}$ of nodes in
  network $i$ depends on nodes in network 3.  Pairs of networks which
  are not connected by dependency links do not have nodes that
  directly depend on each other.
}\label{fig1}
\end{figure}\label{fig1}

\begin{figure}
\centering
\includegraphics[width=0.47\textwidth]{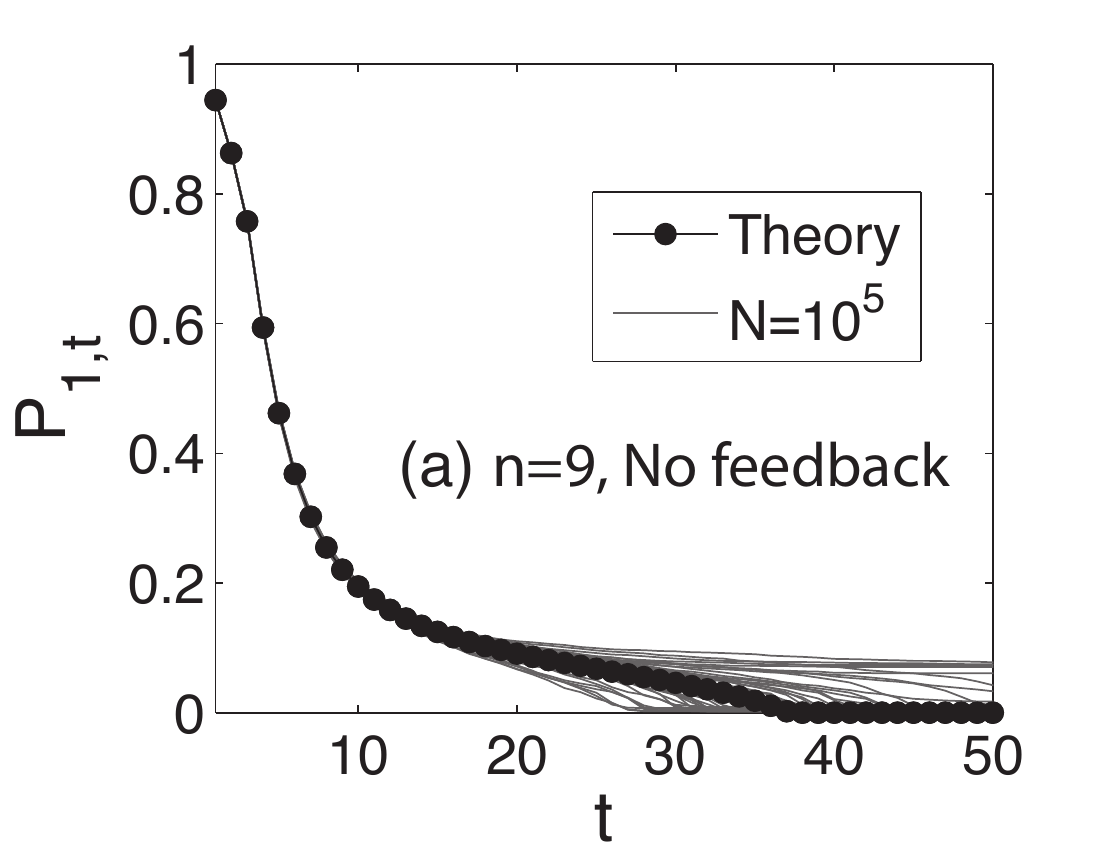}\hspace{.1in}
\includegraphics[width=0.47\textwidth]{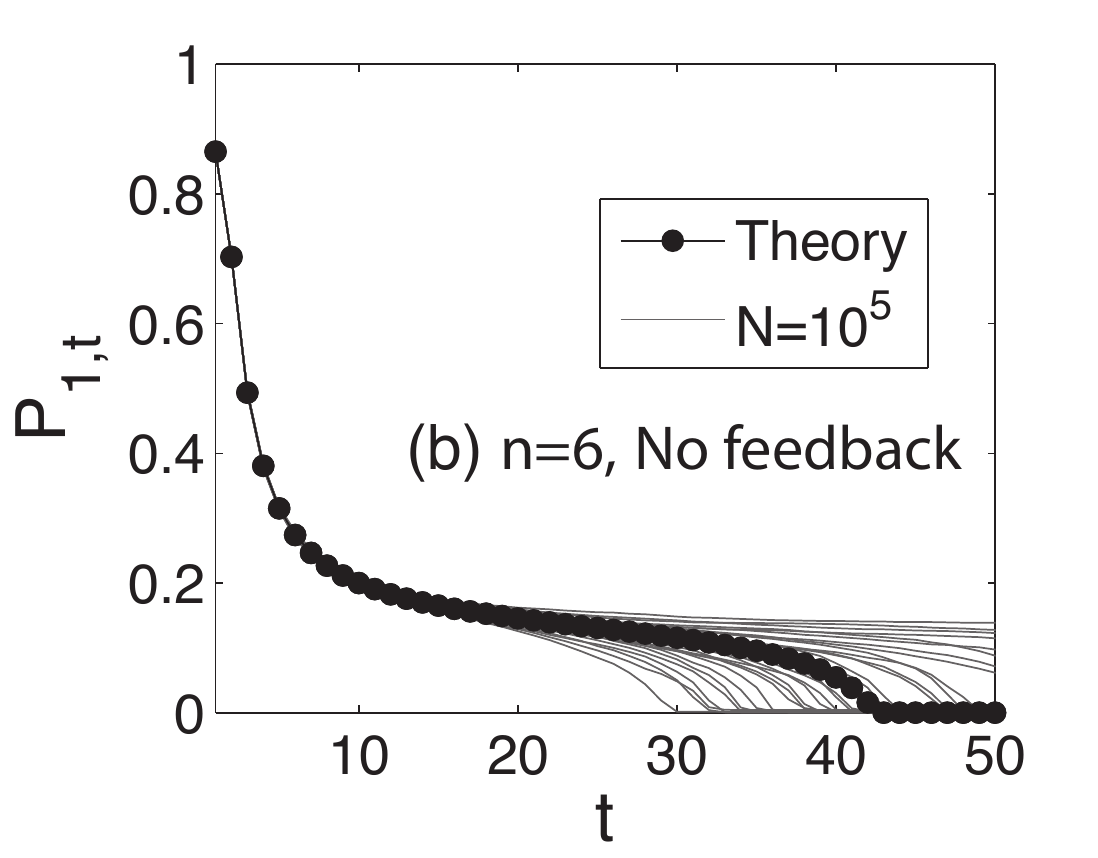}
\includegraphics[width=0.47\textwidth]{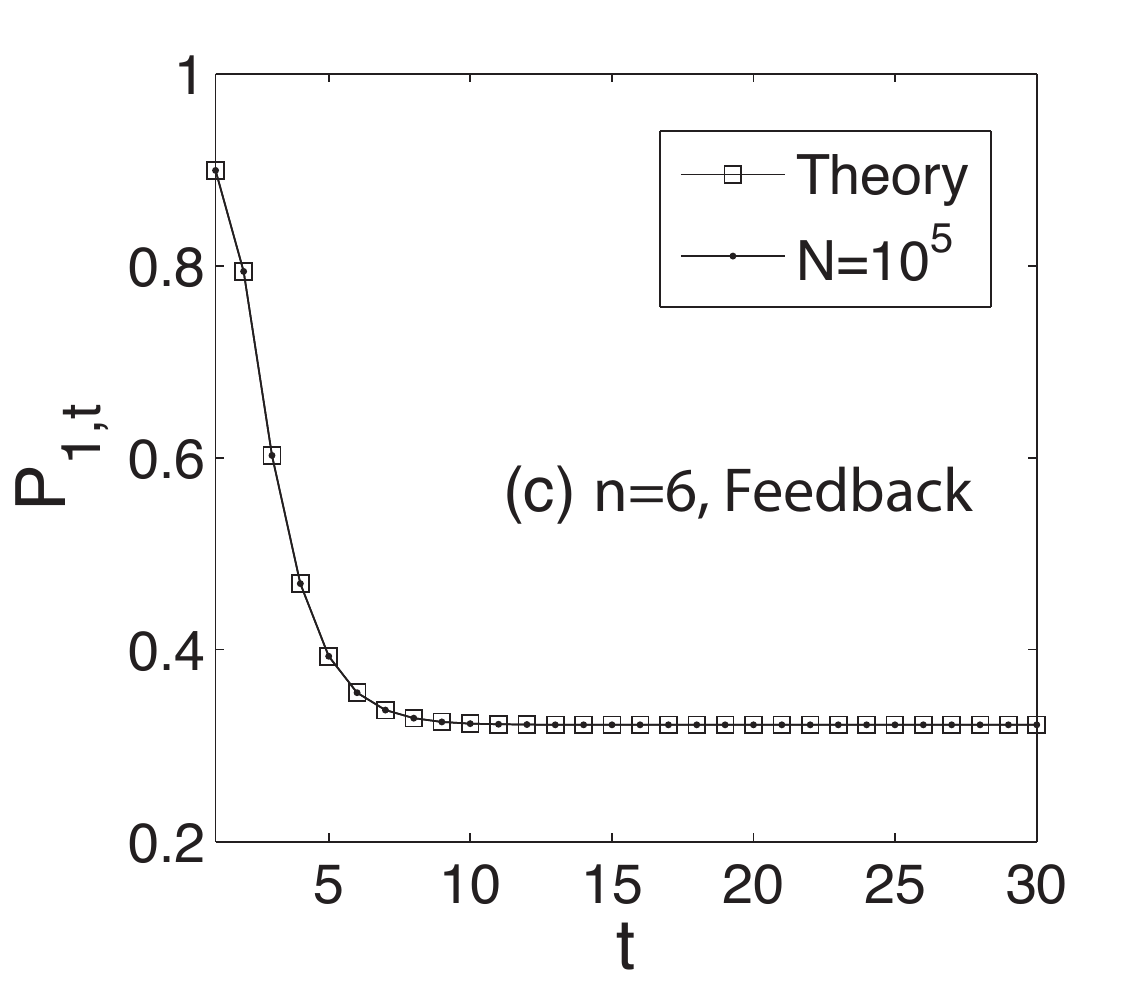}
\caption{(a) Simulation results compared with theory of the giant
component of network $1$, $P_{1,t}$, after $t$ cascading failures
for the lattice NetONet composed of 9 ER networks shown in
Fig.~1(a). For each network in the NetONet, $N=10^5$, $m=4$ and
$\bar{k} = 8$, and $q=0.4>q_c\doteq0.382$ (predicted by Eq.
(\ref{E25}). The chosen value of $p$ is $p=0.945$, and the predicted
threshold is $p^I_c=0.952$ (from Eq. (\ref{GE18}). (b) Simulations
compared to theory of the giant component, $P_{t,1}$, for the random
regular NetONet composed of 6 ER networks shown in Fig.~1(b) with
the no feedback condition. For each network in the NetONet,
$N=10^5$, $m=3$, $\bar{k} = 8$, $q=0.49>q_c\doteq0.4313$ (predicted
by Eq. (\ref{E25})), and for $p=0.866<p^I_c\doteq0.8696$  (from Eq.
(\ref{GE18})). (c) Simulations compared to theory of the giant
component, $P_{1,t}$, for the random regular NetONet composed of 6
ER networks shown in Fig.~1(b) with the feedback condition. For each
network in the NetONet, $N=10^5$, $m=3$ $\bar{k} = 8$,
$q=0.4<q_{\max}=0.5$ (predicted by Eq. (\ref{GE27})), and for
$p=0.9>p_c\doteq0.5787$ (from Eq. \ref{GE25})). The results are
averaged over 20 simulated realizations of the giant component left
after $t$ stages of the cascading failures which is compared with
the theoretical prediction of Eq. (\ref{GE10}).}\label{fig2}
\end{figure}

\begin{figure}[h!]
\centering \includegraphics[width=0.47\textwidth]{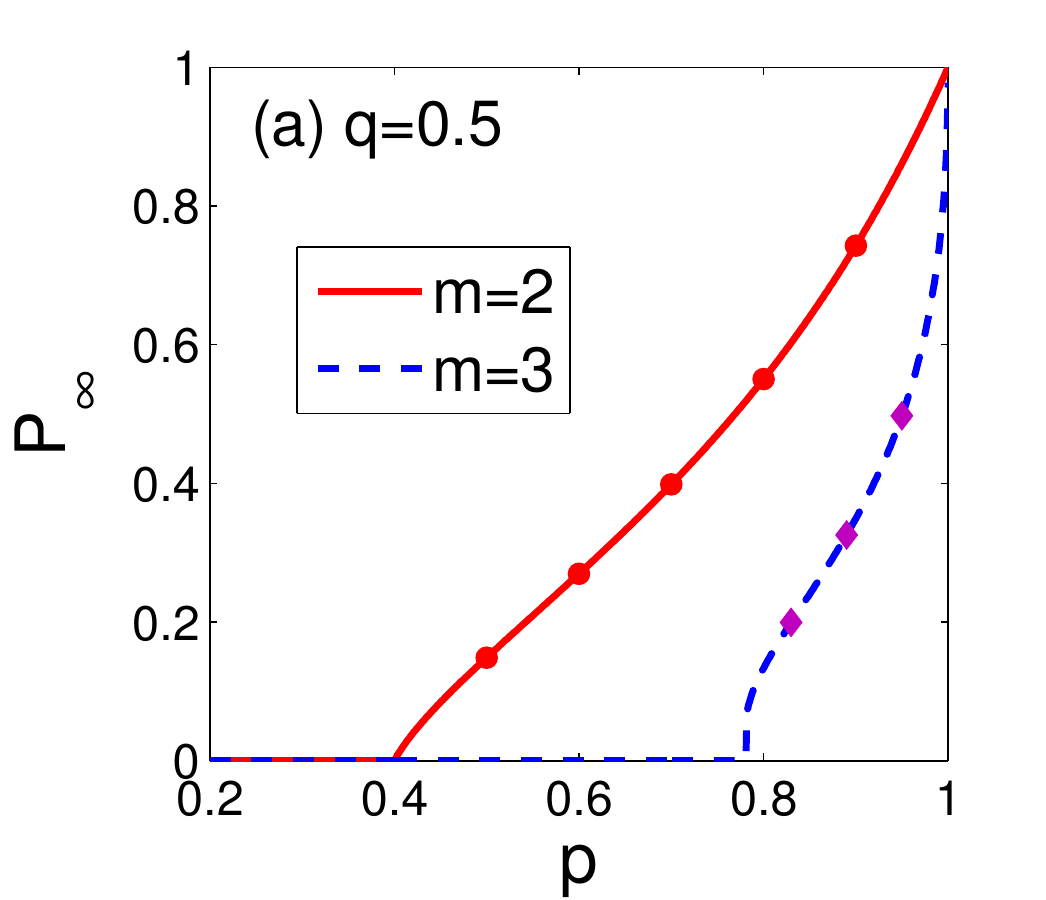}
\hspace{.0in} \includegraphics[width=0.47\textwidth]{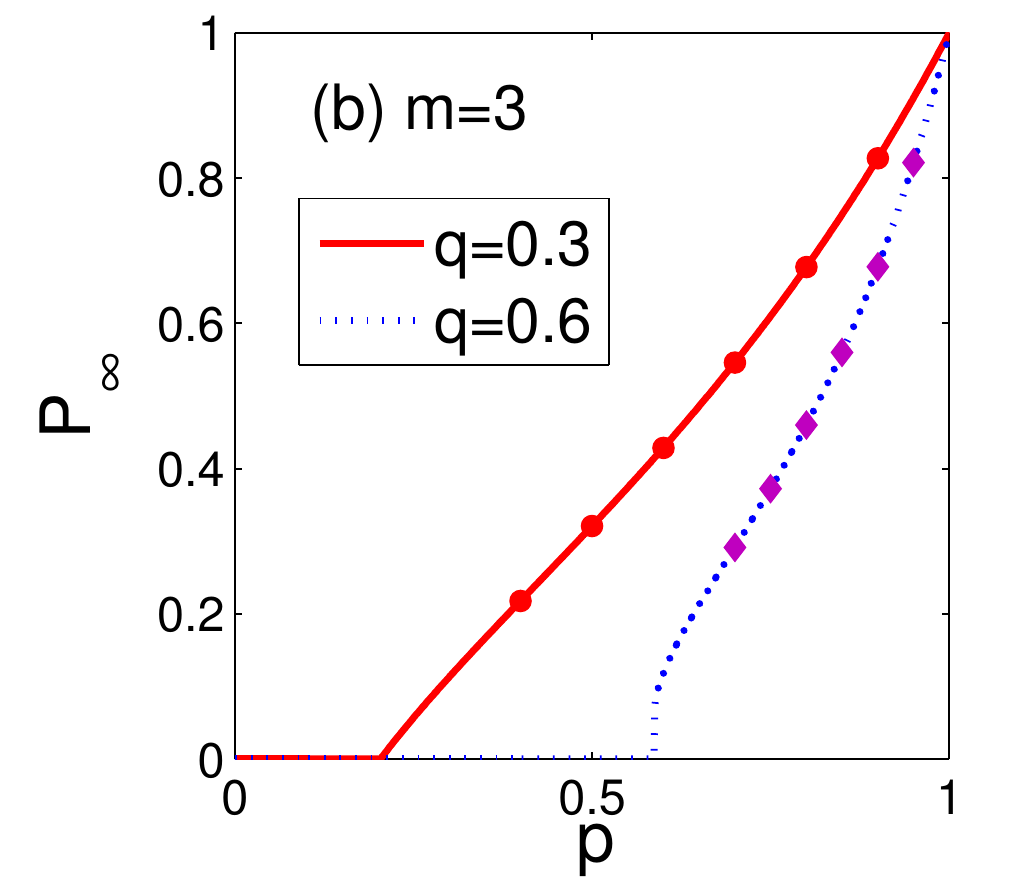}
\caption{ The giant component for an RR network of ER networks,
$P_\infty$, as a function of $p$, for ER networks with average
degree $\bar{k}=10$, (a) for two different values of $m$ and
$q=0.5$, (b) for two different values of $q$ and $m=3$. The curves
in (a) and (b) are obtained using Eq.  (\ref{E16}) and are in
excellent agreement with simulations. The points symbols are
obtained from simulations by averaging over 20 realizations for
$N=2\times 10^5$. In (a), simulation results are shown as circles
($n=6$) for $m=2$ and as diamonds ($n=12$) for $m=3$. These
simulation results support our theoretical result, Eq. (\ref{E16}),
which is indeed independent of number of networks $n$. }\label{fig3}
\end{figure}

\begin{figure}[h!]
\centering
\includegraphics[width=0.49\textwidth]{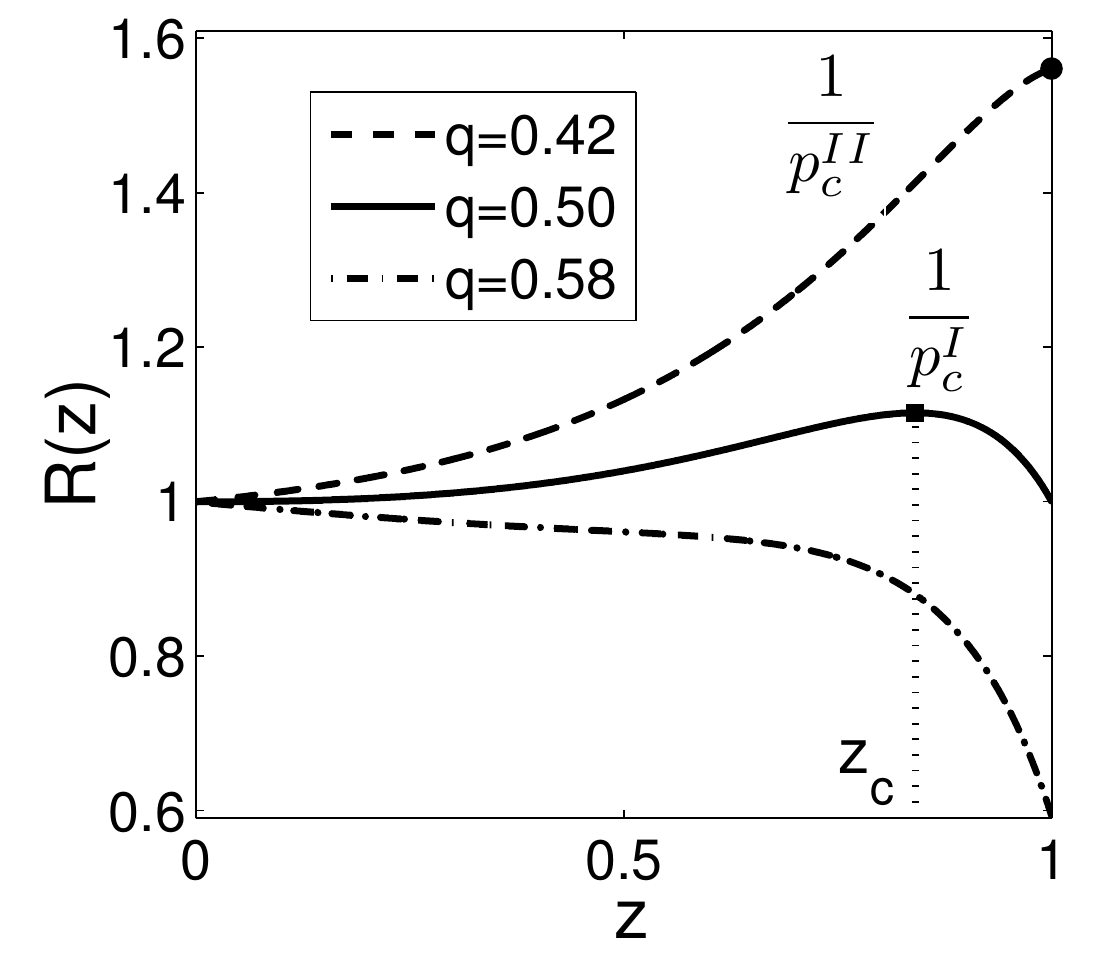}
\caption{Plot of $R(z)$ as a function of $z$ for an RR network of ER
networks, for different values of $q$ when $m=3$ and $\bar{k}=8$.
All the lines are produced using Eq. (\ref{GE15}). The symbols
$\bullet$ and $\blacksquare$ show the critical thresholds $p^{II}_c$
when $q=0.42<q_c=0.4313$ and $p^{I}_c$ when $q=0.5<q_{\max}=0.5462$.
These critical thresholds coincide with the results in Fig.~3(a).
The dashed dotted line shows that when $q=0.58>q_{\max}$ the
function (\ref{GE15}) has no solution for $p=1$, which corresponding
to the case of complete collapse of the NetOnet.}\label{fig4}
\end{figure}

\begin{figure}[h!]
\centering
\includegraphics[width=0.47\textwidth]{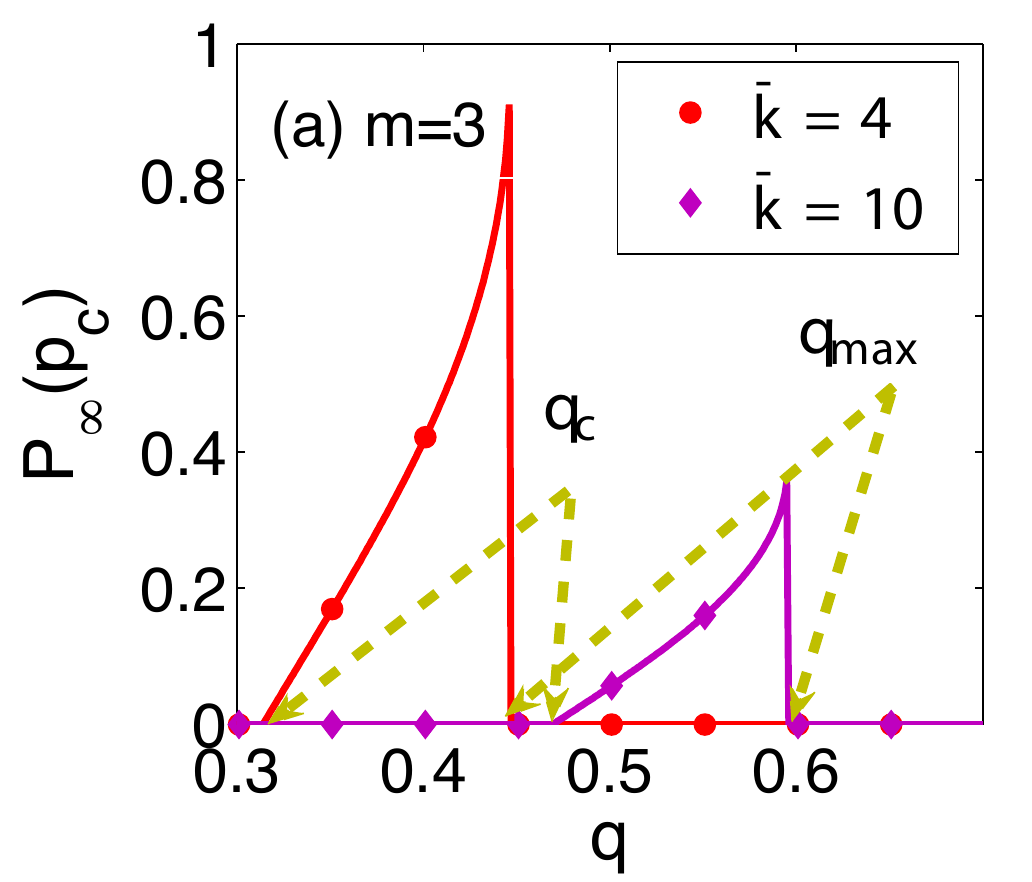} \hspace{.0in}
\includegraphics[width=0.47\textwidth]{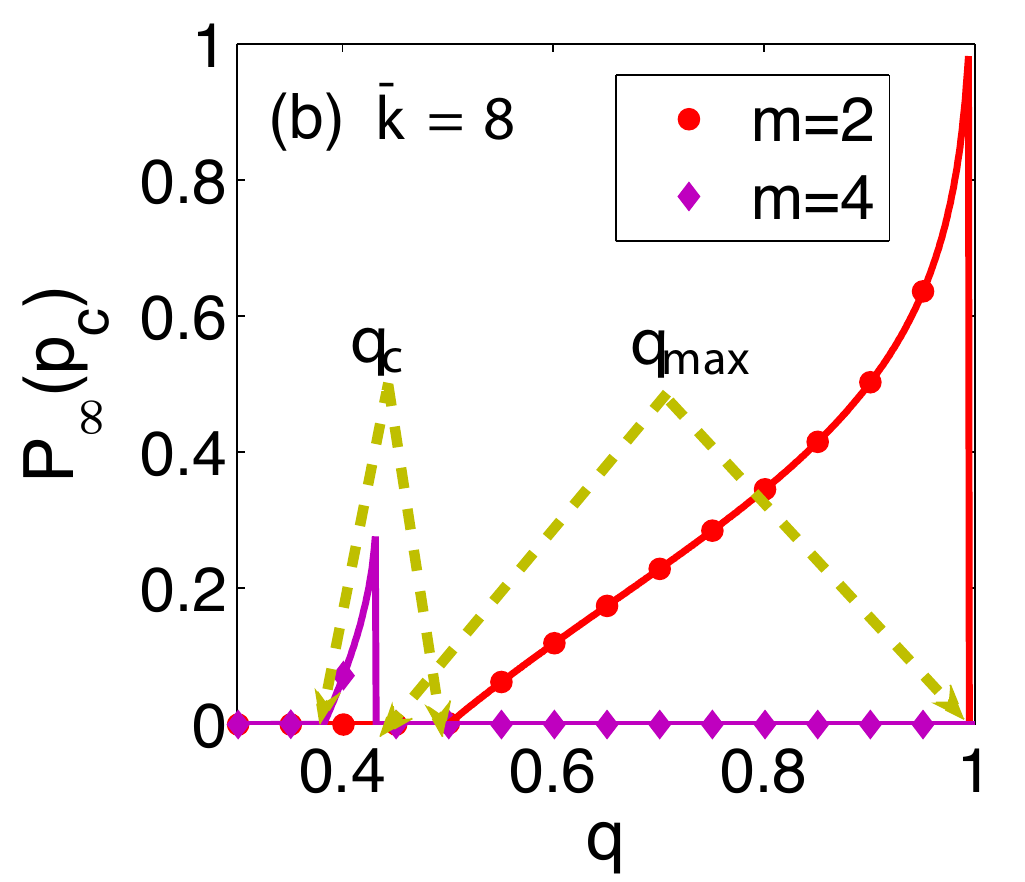} \hspace{.0in}
\caption{ The giant component
  for an RR NetONet formed of ER networks at $p_c$, $P_\infty(p_c)$, as a
  function of $q$. The curves are (a)
  for $m=3$ and two different values of $\bar{k}$, and (b) for $\bar{k}=8$ and two different values of $m$.
  The curves are obtained using
  Eqs. (\ref{E16}) and (\ref{E20}) and are in excellent agreement with
  simulations (symbols). Panels (a) and (b) show the location of $q_{\max}$ and $q_c$ for
  two values of $m$. Between $q_c$ and $q_{max}$ the transition is first
  order represented by $P_{\infty}(p_c)>0$. For $q<q_c$ the transition
  is second order since $P_{\infty}(p_c)=0$ and for $q>q_{max}$ the
  NetONet collapses ($P_{\infty}(p_c)=0$) and there is no phase transition ($p_c=1$).}\label{fig5}
\end{figure}

\begin{figure}[h!]
\centering
\includegraphics[width=0.47\textwidth]{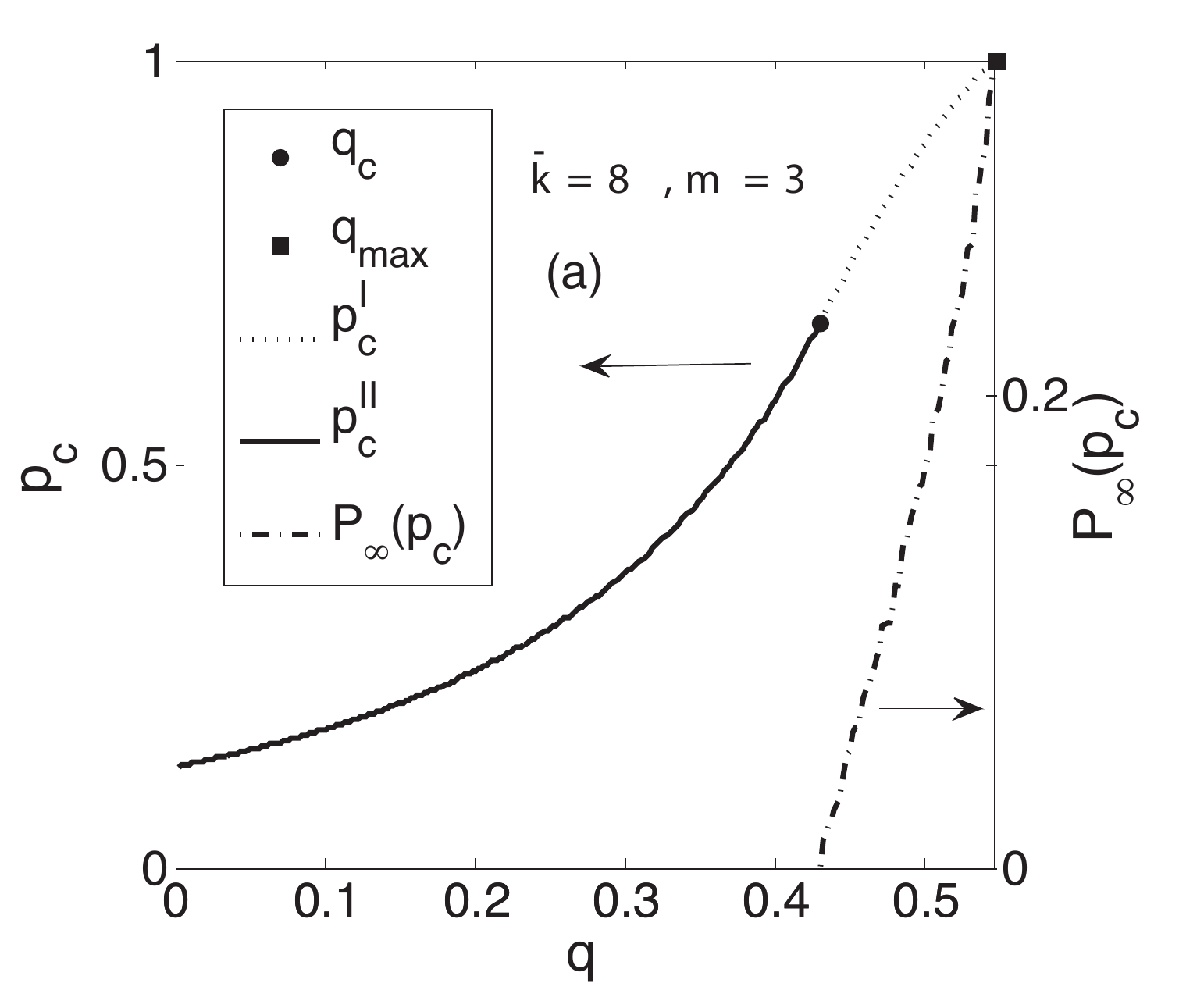} \hspace{.0in}
\includegraphics[width=0.47\textwidth]{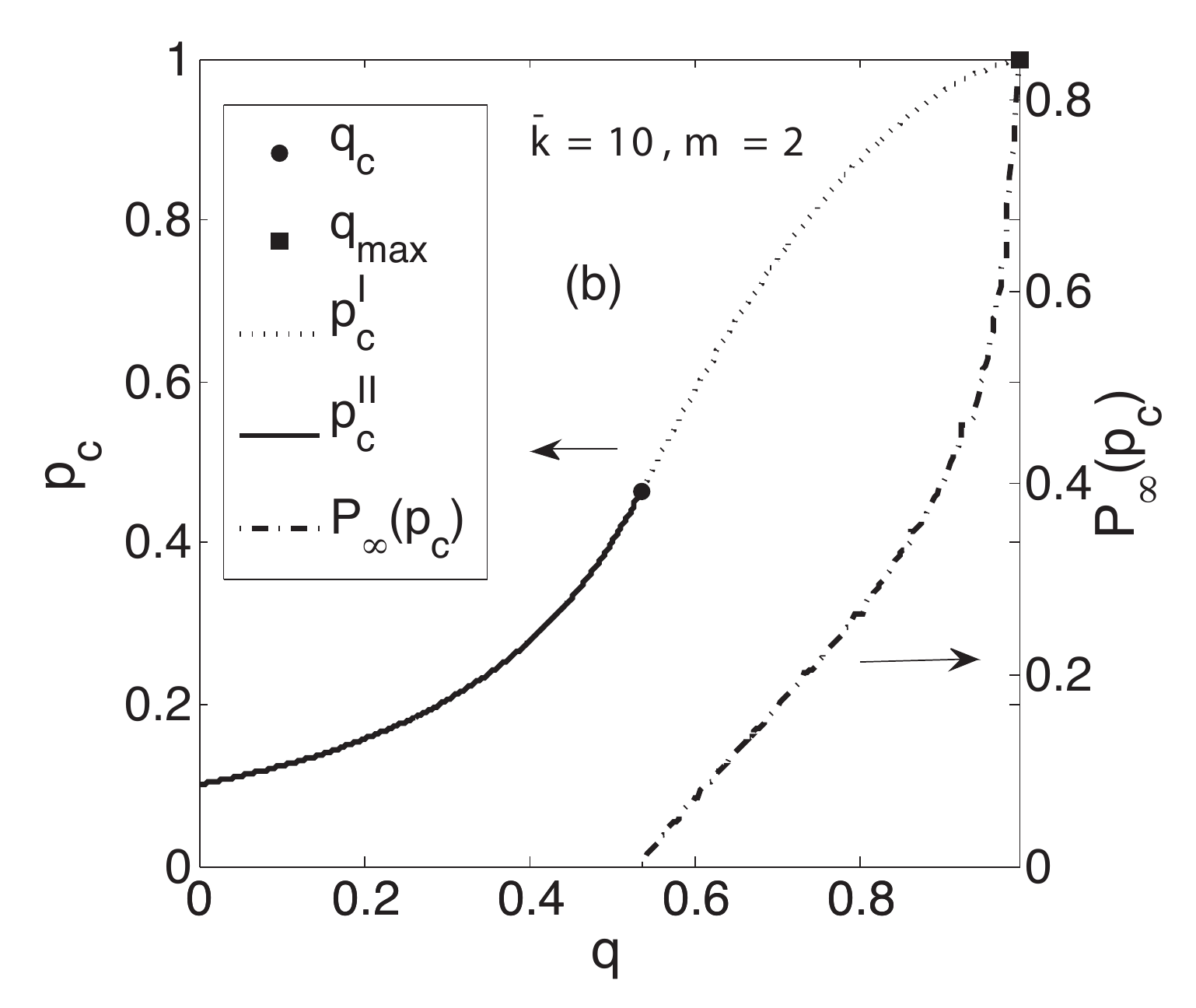} \hspace{.0in}
\caption{The phase diagram for RR network of ER networks, (a) for
$m=3$ and $\bar{k}=8$, (b) for $m=2$ and $\bar{k}=10$. The solid
curves show the second order phase transition (predicted by Eq.
(\ref{E25})) and the dashed-dotted curves show the first order phase phase
transition, leading $P_\infty(p_c)$ at $q_c$ from zero to non-zero
values (the rhs axis). As $m$ decreases and $\bar(k)$ increases,
the region for $P_{\infty}>0$ increases, showing a better
robustness. The circle shows the tri-critical point $q_c$, below
which second order transition occurs and above which a first order
transition occurs. The square shows the critical point $q_{\max}$,
above which the NetONet completely collapse even when
$p=1$.}\label{fig5-1}
\end{figure}


\begin{figure}[h!]
\centering
\includegraphics[width=0.47\textwidth]{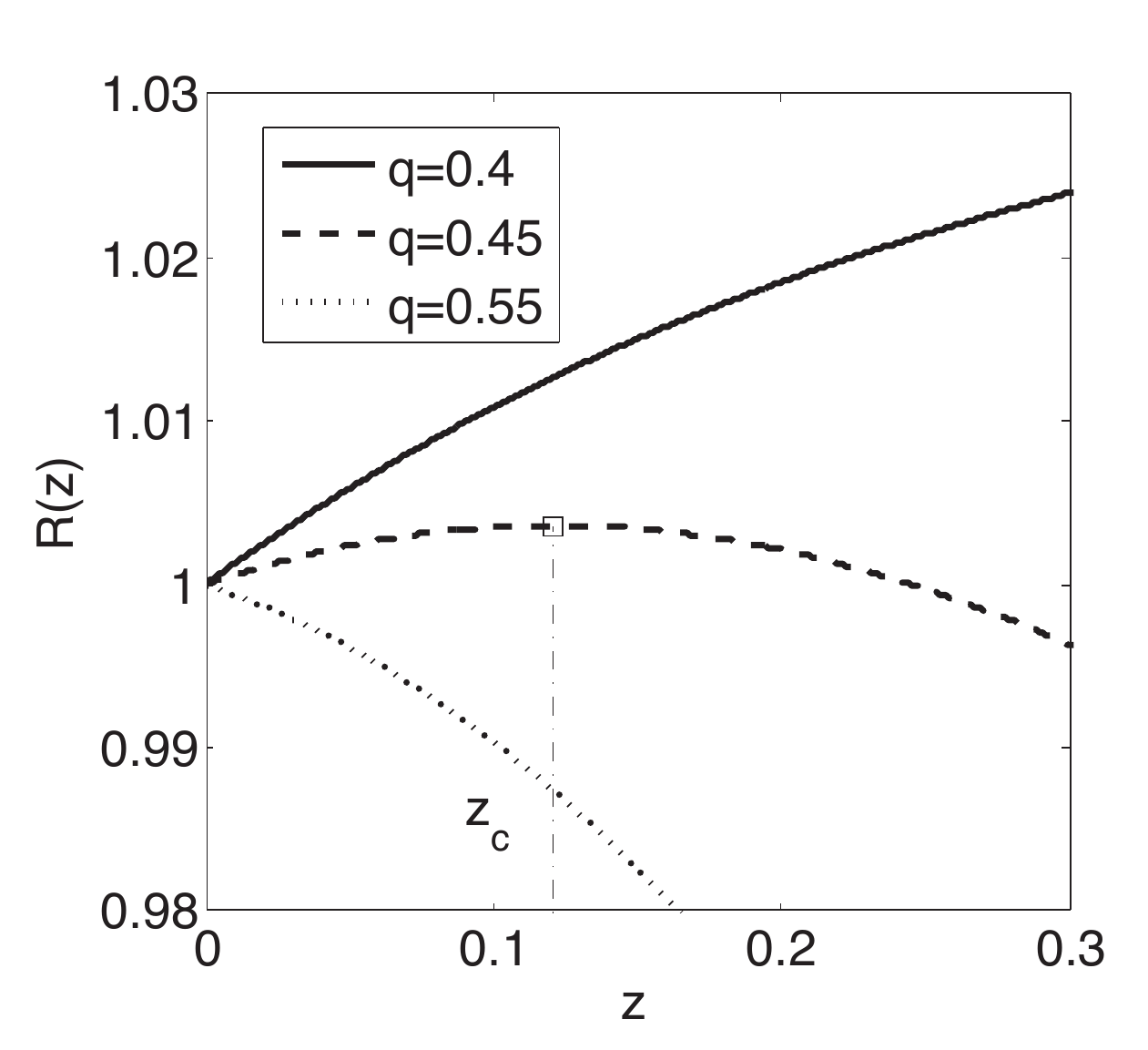} \hspace{.0in}
\caption{For RR NetONet formed of SF networks $R(z)$ as a function of $z$
for different values of $q$ when $m=3$, $\lambda=2.3$, $s=2$ and
$M=1000$. (i) When $q$ is small ($q=0.4<q_c$), $R(z)$ is a
monotonically increasing function of $z$, the system shows a second
order phase transition. (ii) When $q$ is larger
($q_c<q=0.45<q_{\max}$), $R(z)$ as a function of $z$ shows a peak at
$z_c$ which corresponds to a hybrid phase transition. The square
symbol represents the critical point of the sharp jump ($z_c$). (iii) When
$q$ is large enough ($q=0.55>q_{\max}$), $R(z)$ decreases with $z$
first, and then increases with $z$, which corresponds the system
collapses.}\label{fig5-20}
\end{figure}

\begin{figure}[h!]
\centering
\includegraphics[width=0.4\textwidth]{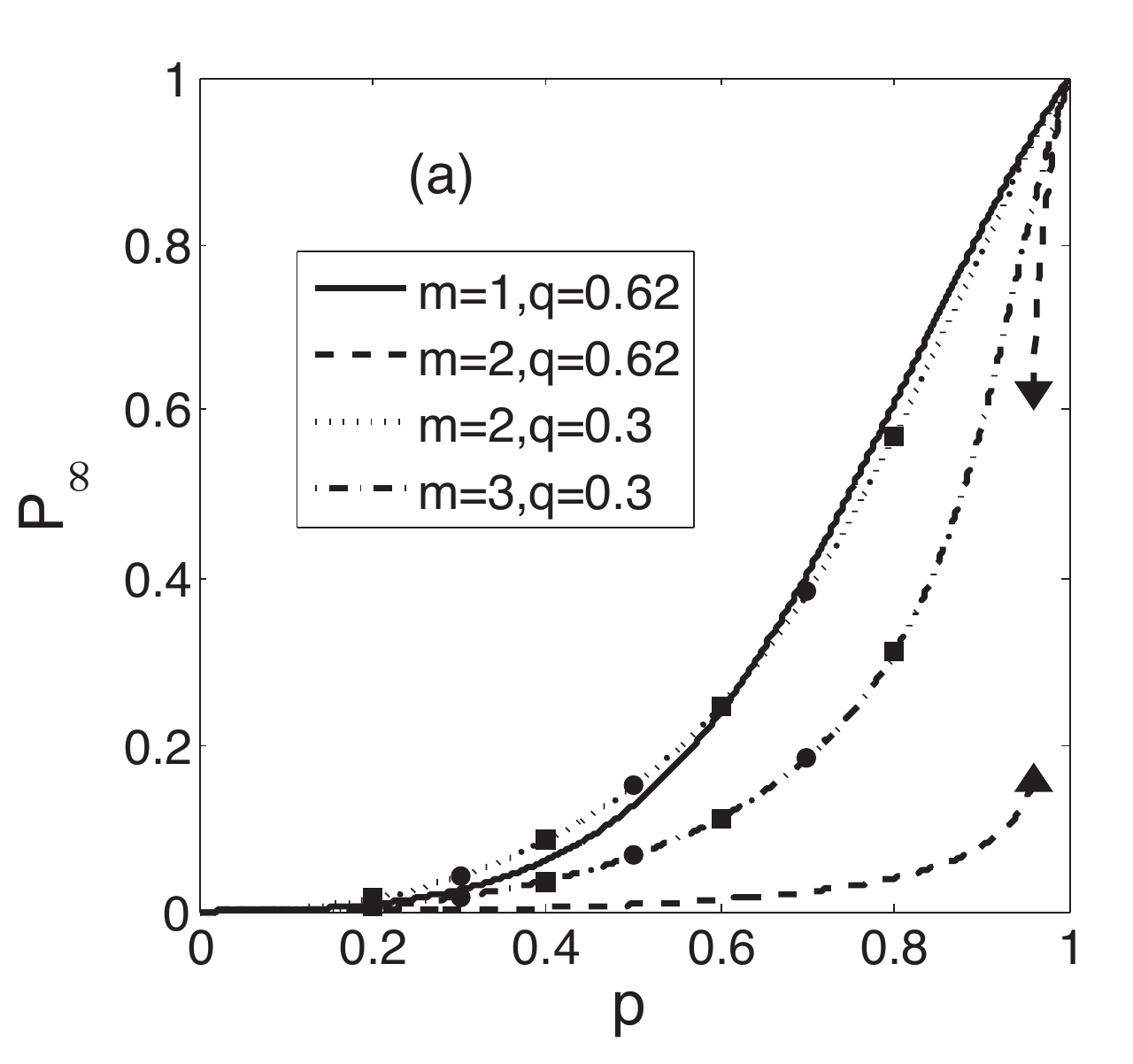} \hspace{.0in}
\includegraphics[width=0.47\textwidth]{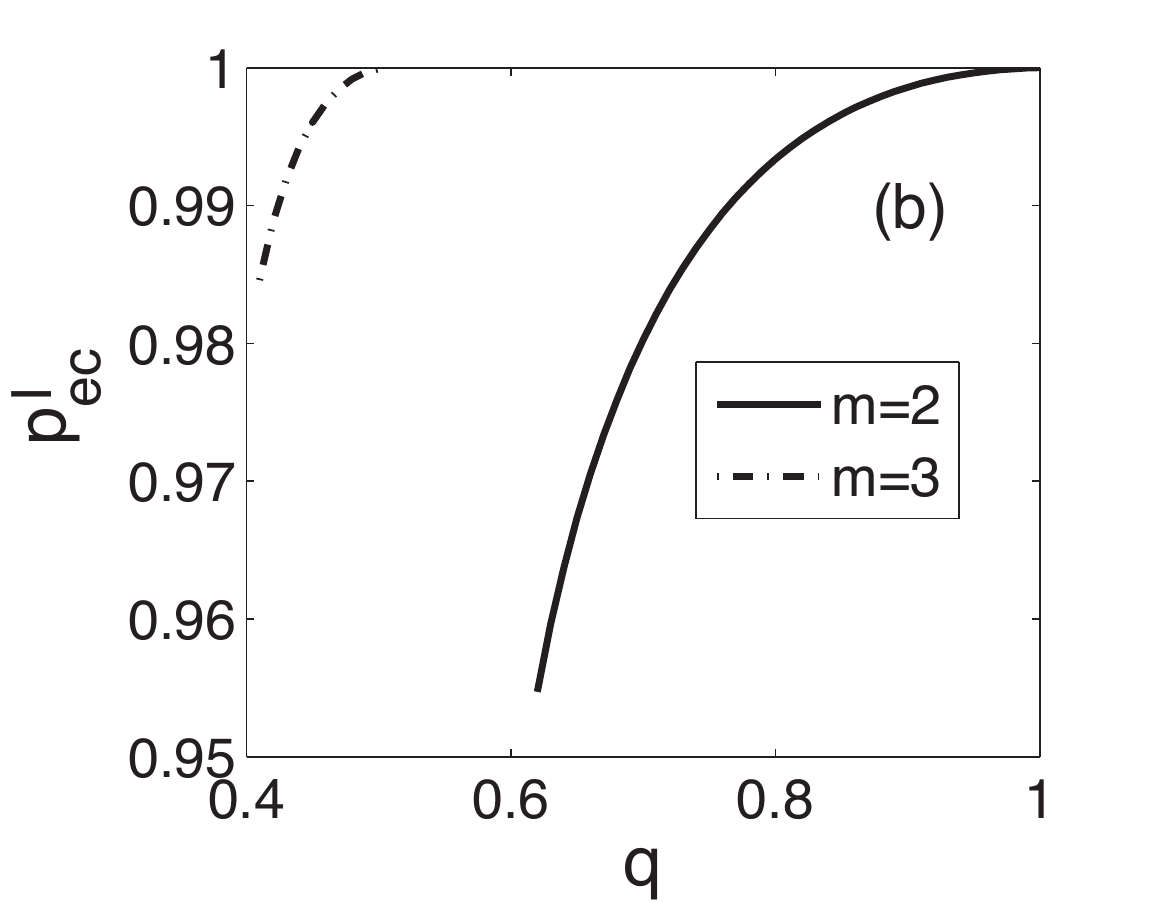} \hspace{.0in}
\includegraphics[width=0.47\textwidth]{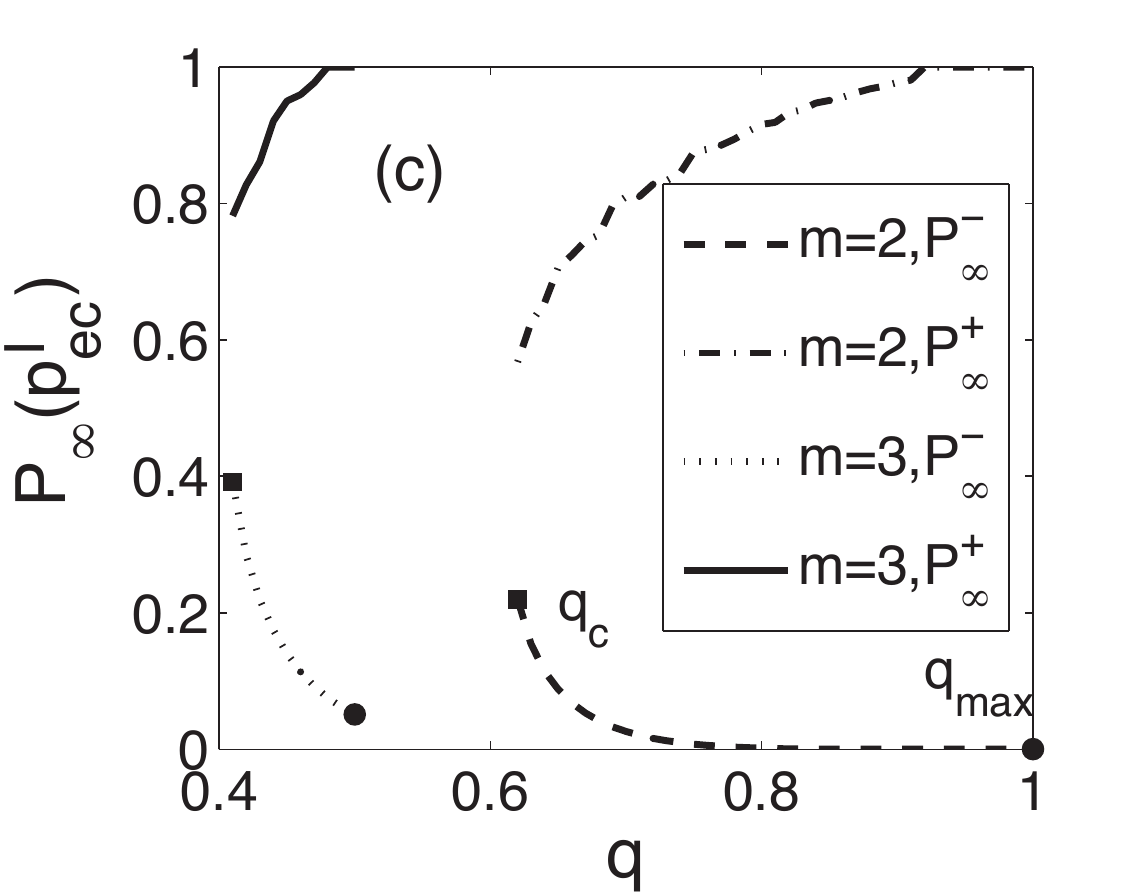} \hspace{.0in}
\caption{Results for a RR network formed of SF networks. (a) The giant
component $P_{\infty}$ as a function of $p$ for different values of
$m$ and $q$ for $\lambda=2.5$. (b) The critical threshold $p^I_{ec}$
and (c) the corresponding giant component at the threshold
$P_{\infty}(p^I_{ec})$ as a function of coupling strength $q$ for
$m=2$ and $m=3$. The symbols in (a) represent simulation results,
obtained by averaging over 20 realizations for $N=2\times10^5$ and
number of networks $n=6$ (squares) and $n=4$ (circles). The lines
are the theoretical results obtained using Eqs. (\ref{GE5}) and
(\ref{GE1})-(\ref{GE3}). We can see in (a) that the system shows a
hybrid phase transition for $m=2$ and
$q^e_c<q=0.62<q_{\max}=1/(m-1)$. When $q<q^e_c$ the system shows a
second order phase transition and the critical threshold is
$p^{II}_c=0$. However, in the simulation when $p$ is small (but not zero)
$P_{\infty}=0$. This happens because $p^{II}_c=0$ is valid only when
the network size $N=\infty$ and $M=\infty$, but in simulations we have finite systems. Furthermore, when
$q^e_c<q<q_{\max}$ the system shows a hybrid transition shown in (a)
and (c), and when $q>q_{\max}$ all the networks collapse even if one
node fails. We call this hybrid transition because $P^-_{\infty}>0$,
which is different from the case of ER networks with first order
phase transition where $P^-_{\infty}=0$.}\label{fig5-21}
\end{figure}

\begin{figure}[h!]
\centering \includegraphics[width=0.47\textwidth]{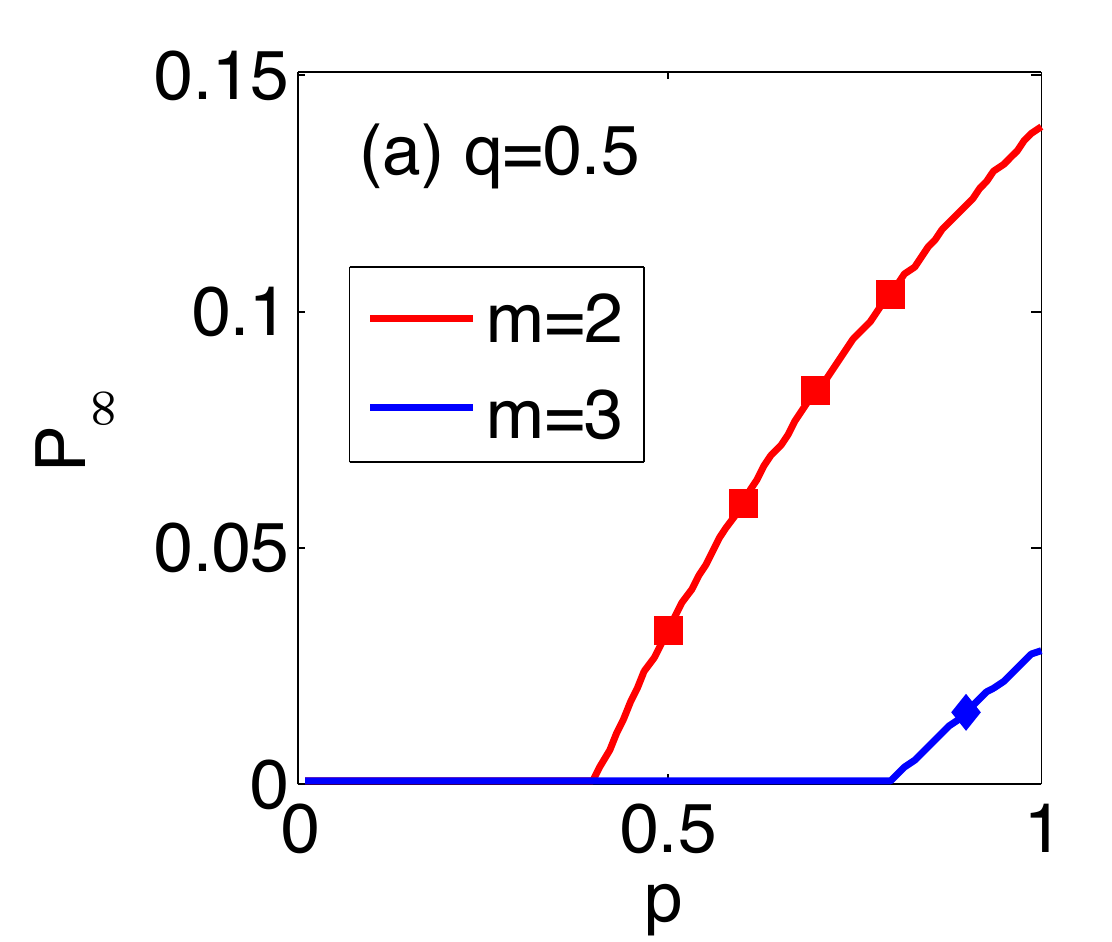}
\centering \includegraphics[width=0.47\textwidth]{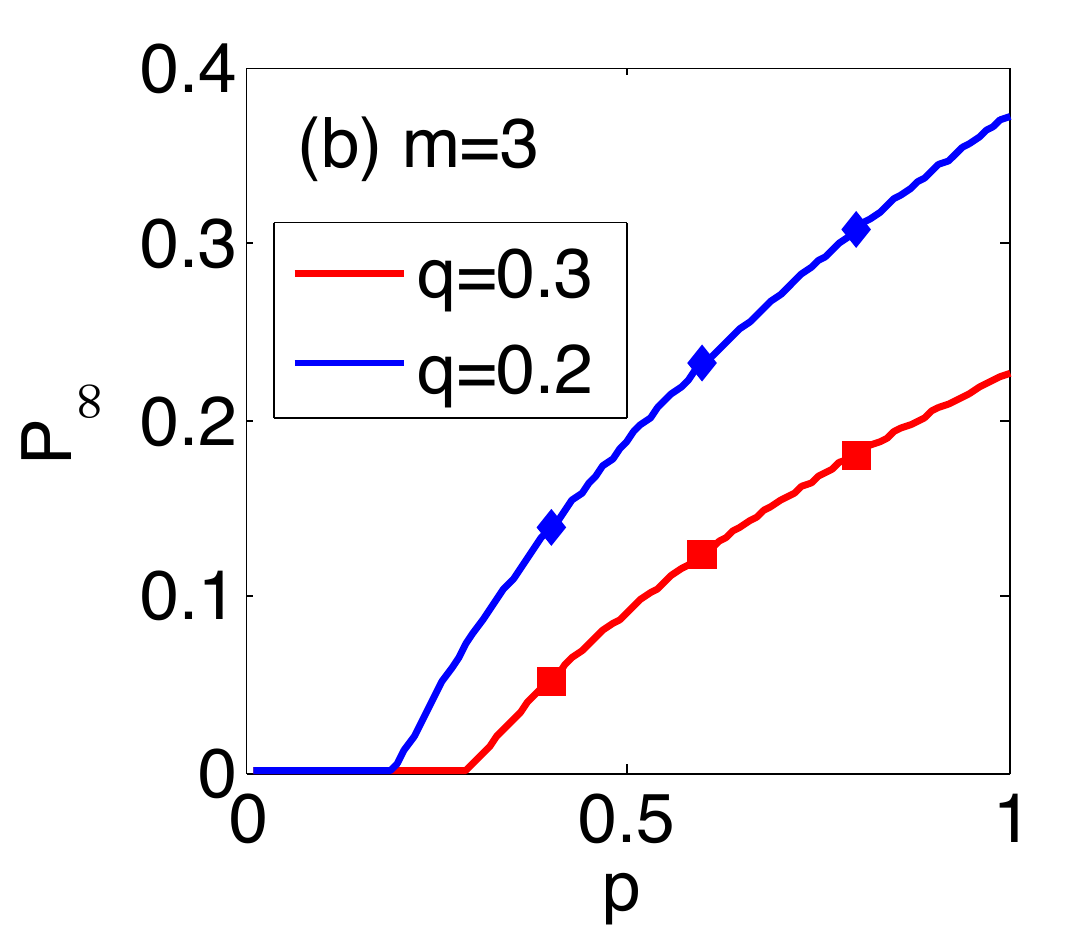}
\caption{The giant component for an RR network of ER networks with
feedback condition, $P_\infty$, as a function of $p$ for ER average
degree $\bar{k}=10$, for different values of $m$ when $q=0.5$ (a)
and for different values of $q$ when $m=3$. The curves in (a) and
(b) are obtained using Eq. (\ref{GE22}) and are in excellent
agreement with simulations. The points symbols are obtained from
simulations of Fig. 1(b) topology when $m=3$ and $n=6$ networks
forming a circle when $m=2$ by averaging over 20 realizations for
$N=2\times 10^5$. The absence of first order regime in 
NetONet formed of ER networks is due to the fact that at the initial stage nodes in each network are interdependent on isolated nodes (or clusters) in the other network. However, if only nodes in the giant components of both networks are interdependent, all three regimes, second order, first order and collapse will occur, like in the case of RR NetONet formed of RR networks (see Eq. (50) and Fig. 12). }\label{fig7}
\end{figure}

\begin{figure}[h!]
\centering
\includegraphics[width=0.49\textwidth]{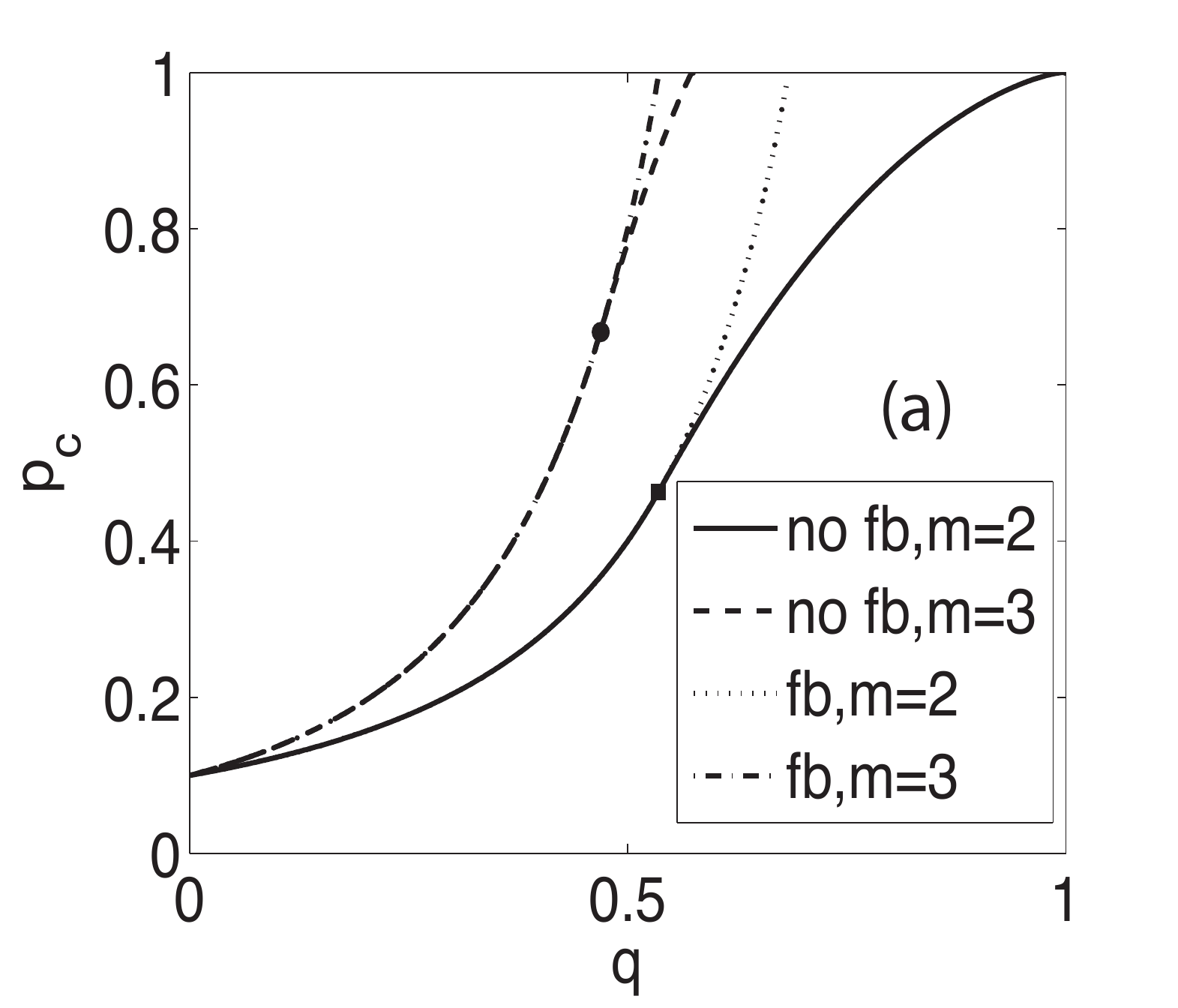}
\includegraphics[width=0.46\textwidth]{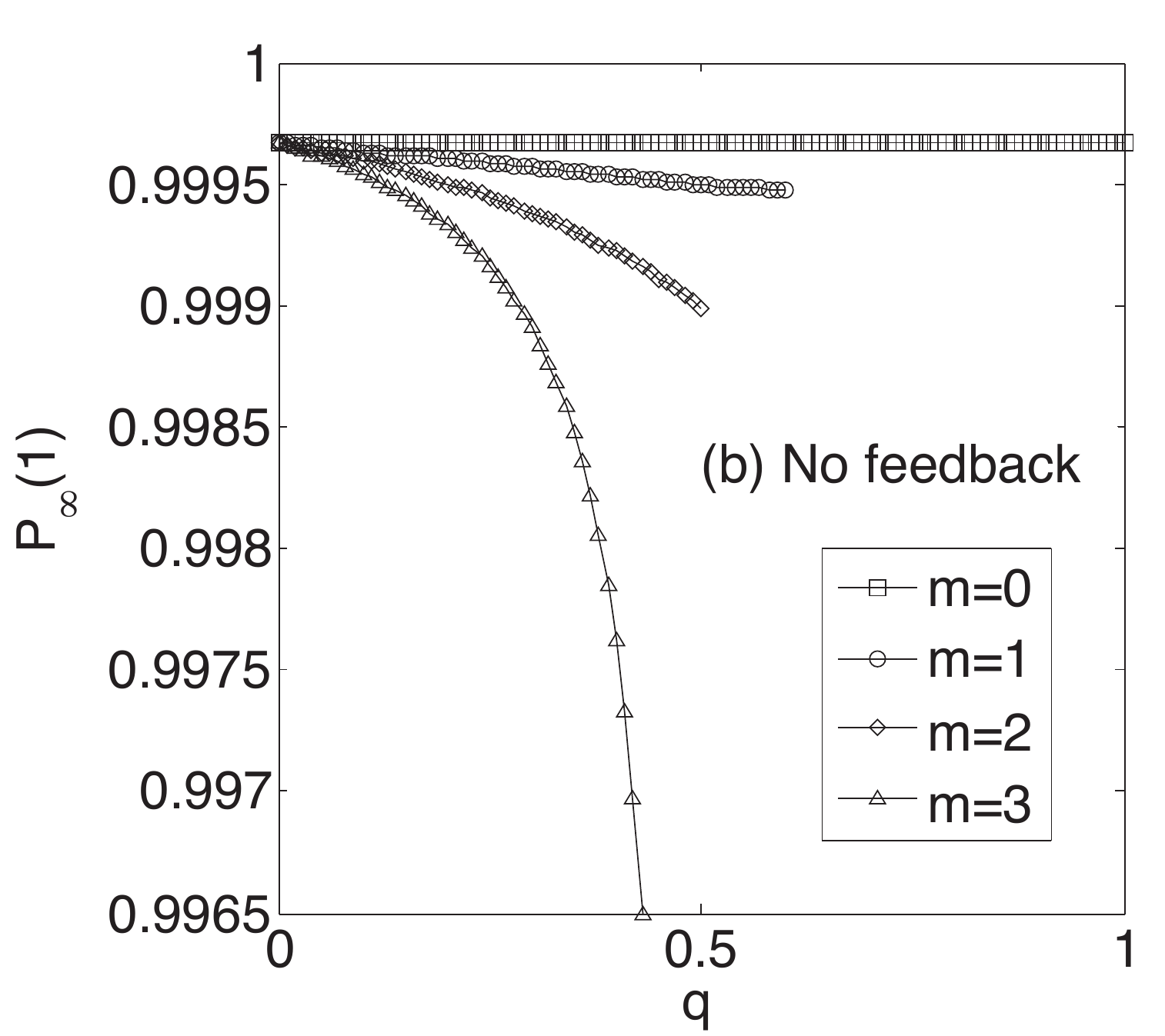}
\includegraphics[width=0.46\textwidth]{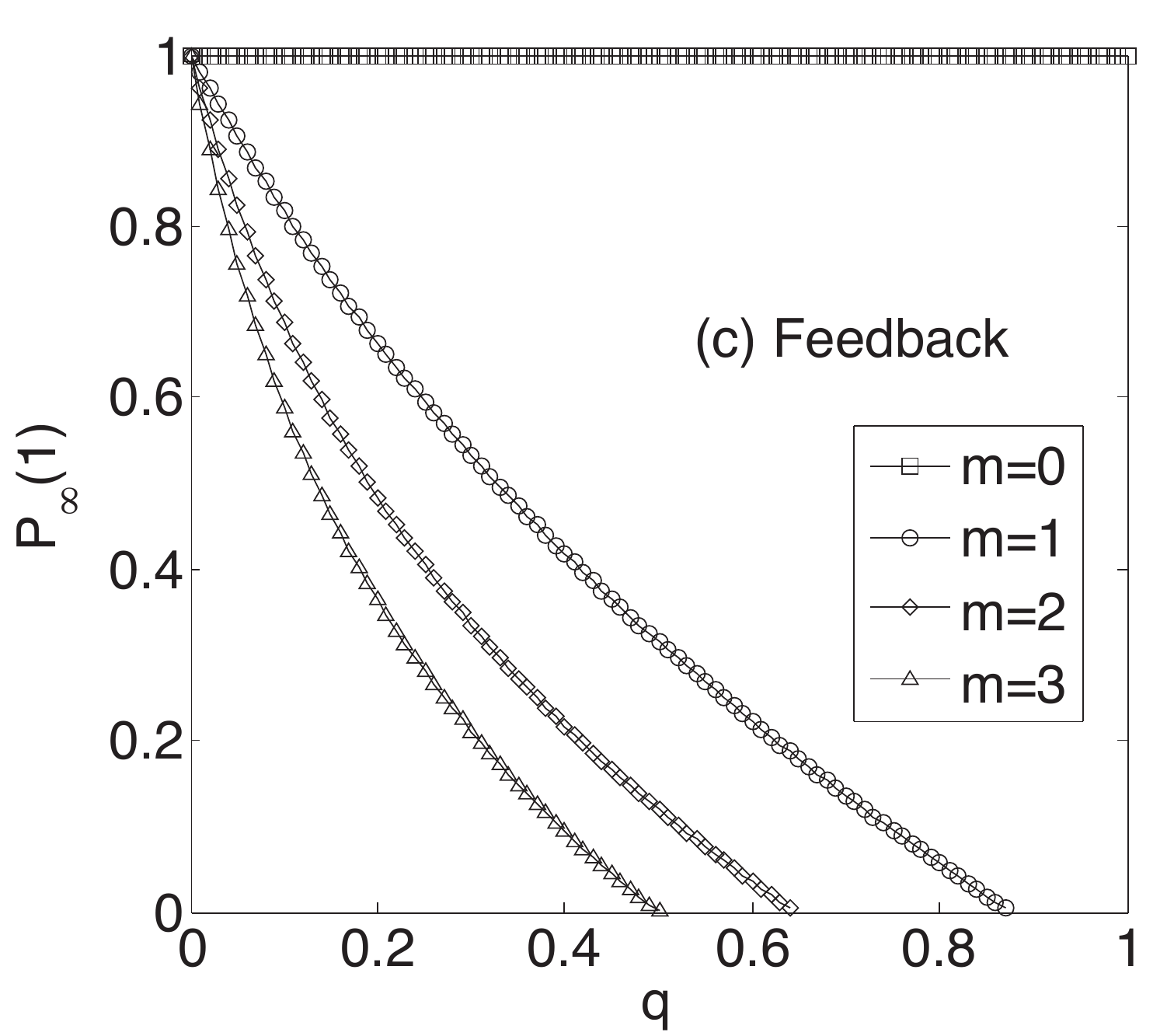}
\caption{(a) $p_c$ as a function of $q$ for both no feedback
condition and feedback condition when $\bar{k}=10$. For no feedback
condition, the parts of curves bellow the symbols show $p^{II}_c$
and above the symbols show $p^{I}_c$. For the feedback condition,
they only have the $p_c$ of second order, and $p^{II}_c$ for the no
feedback case is equal to $p_c$ of the feedback case, but this does
not mean that these two cases have equal vulnerability.
$P_{\infty}(1)$ as a function of $q$ for different values of $m$
when $\bar{k}=8$ with (b) no feedback condition and (c) feedback
condition. When $q=0$, $P_{\infty}(1)=1-exp(-\bar{k})$ for all
$m$ and both feedback and no feedback cases. Comparing (b) and (c),
we can see that the feedback case is much more vulnerable than the
no feedback condition, because $P_{\infty}(1)$ of no feedback case
is much less than that of feedback case.}\label{fig9}
\end{figure}

\begin{figure}[h!]
\centering
\includegraphics[width=0.49\textwidth]{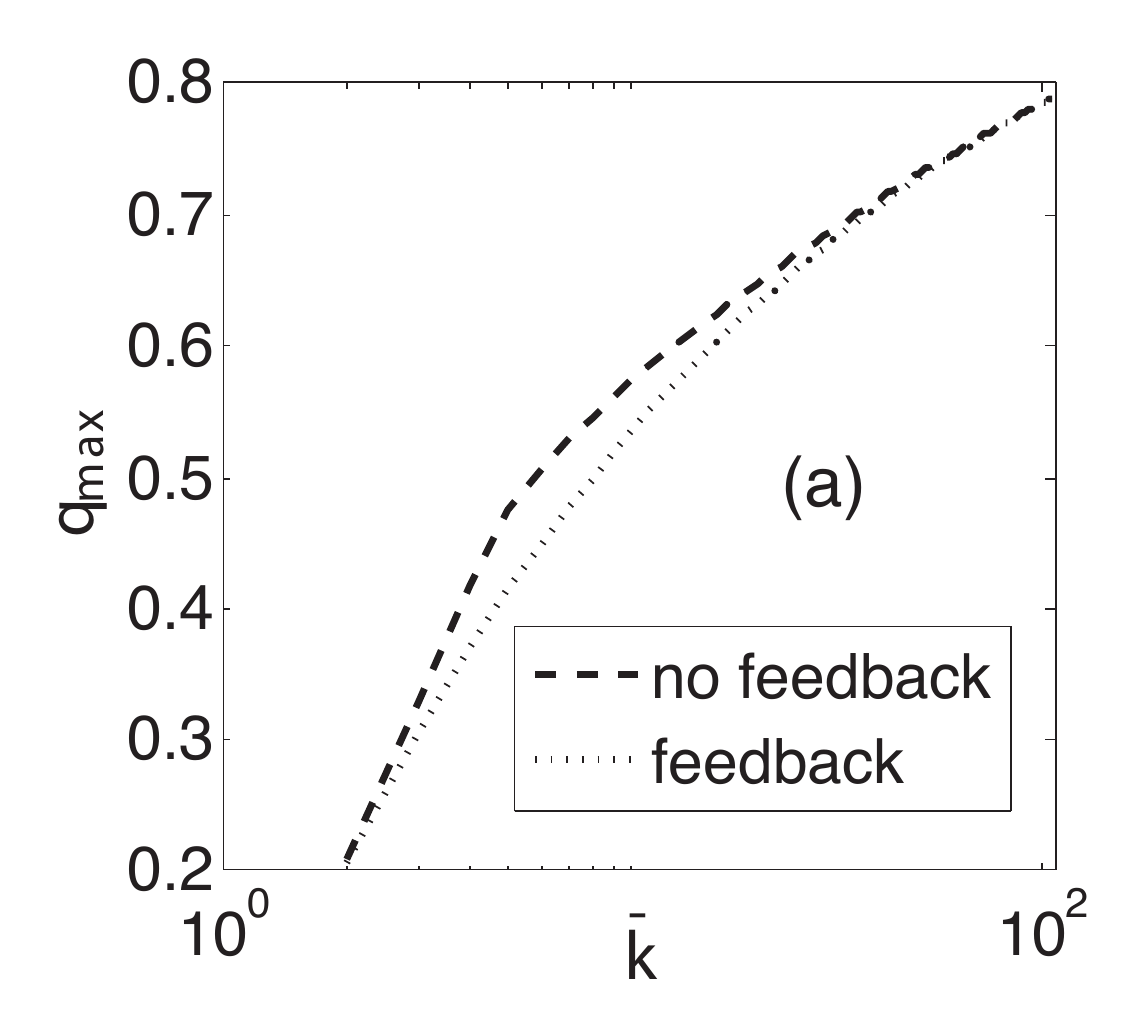}
\includegraphics[width=0.49\textwidth]{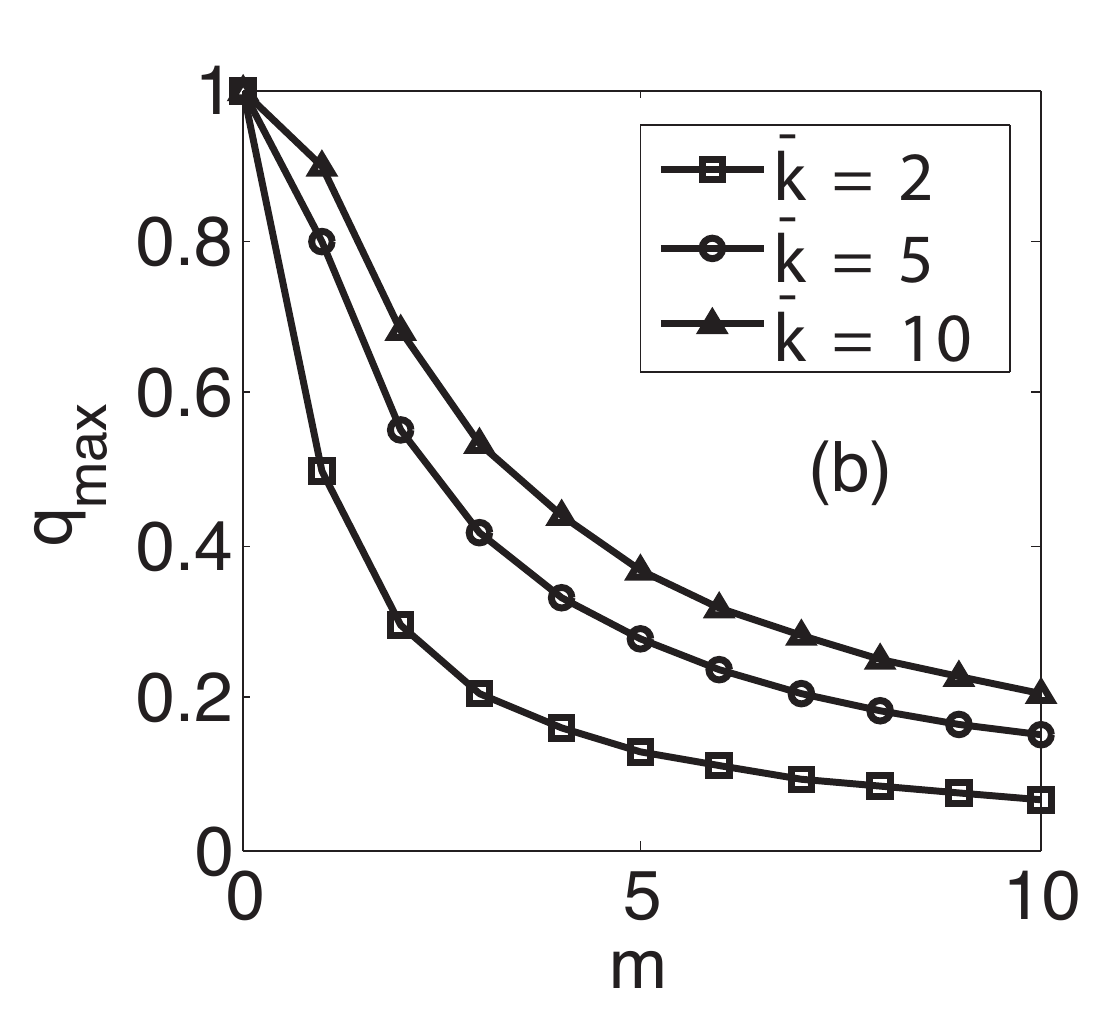}
\caption{(a) The maximum value of coupling strength $q_{\max}$ as a
function of $\bar{k}$ for the case of feedback condition and no
feedback condition when $m=3$.  We can see that $q_{\max}$ of no feedback case is larger than that of the feedback case, which indicate that the no feedback case is more robust compared to the feedback case. (b) The maximum value of coupling strength $q_{\max}$ as a
function of $m$ with the feedback condition for different values of $\bar{k}$, which shows that increasing $\bar{k}$ or decreasing $m$ will increase $q_{\max}$, i.e., increase the robustness of NetONet.}\label{fig8}
\end{figure}

\begin{figure}[h!]
\centering \includegraphics[width=0.47\textwidth]{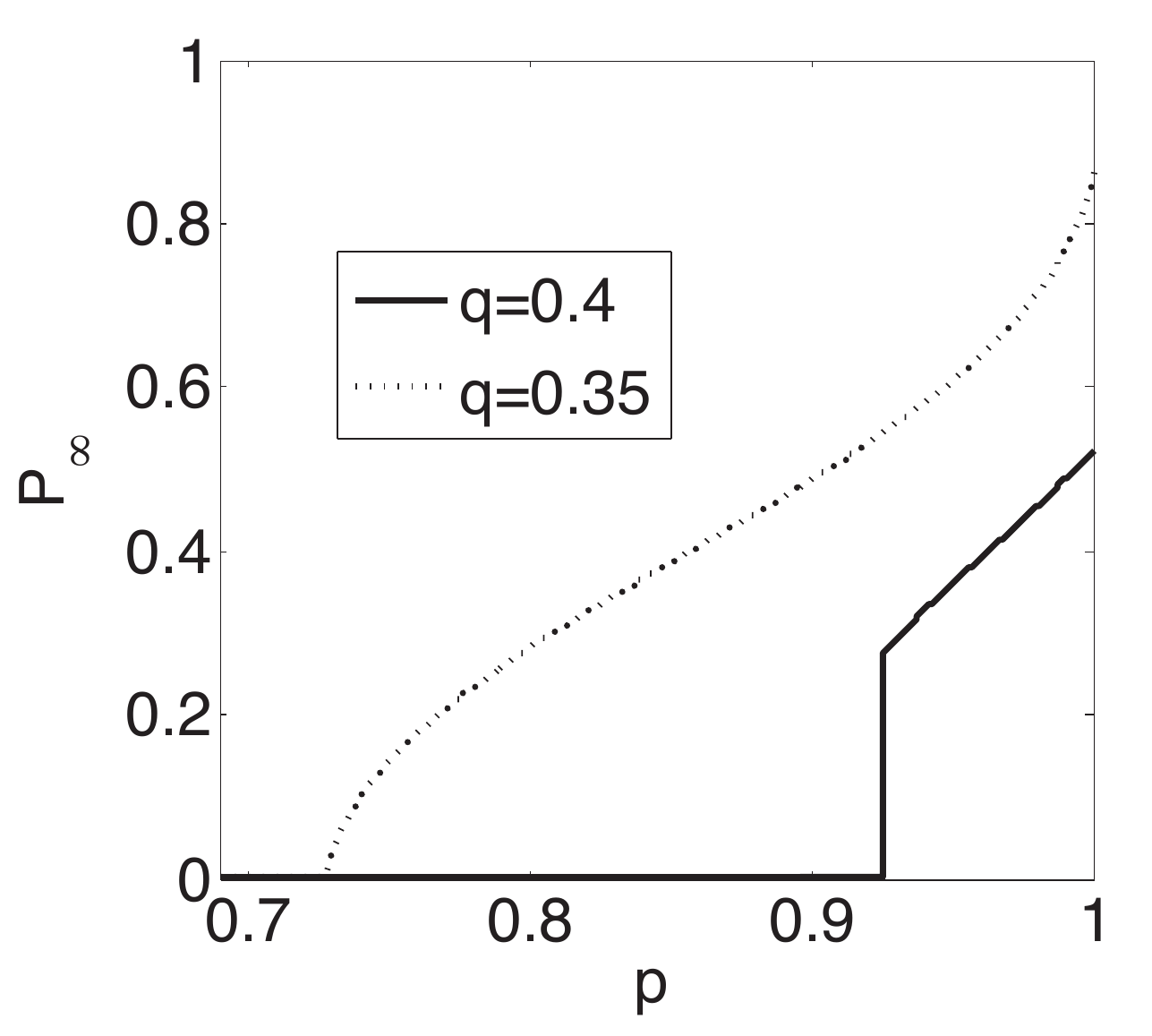}
\caption{The giant component for an RR NetONet formed of RR networks with
feedback condition, $P_\infty$, as a function of $p$ for RR of
degree $k=6$ and $m=3$, for two different values of $q$. The curves are obtained using Eq. (\ref{GE28}), which shows a first order phase transition when $q$ is large but a second order phase transition when $q$ is small.}\label{fig12}
\end{figure}

\end{document}